\documentclass[trackchanges,twocolumn]{aastex701}

\usepackage{comment}
\usepackage{natbib}
\usepackage{orcidlink}
\usepackage{amsmath}

\makeatletter
\newcommand{\noopsort}[1]{}
\makeatother

\begin{document}

\title{Three-Dimensional Ocean Dynamics and Detectability of Tidally Locked Lava Worlds}

\author[0000-0001-9700-9121]{Yanhong Lai}
\affiliation{Tsung-Dao Lee Institute, Shanghai Jiao Tong University, 1 Lisuo Road, Shanghai 200127, China}
\email[show]{yanhonglai@sjtu.edu.cn}  

\author[0000-0002-4615-3702]{Wanying Kang}
\affiliation{Department of Earth, Atmosphere and Planetary Science, MIT, Cambridge, MA 02139, USA}
\email[show]{wanying@mit.edu}

\author[0000-0001-6031-2485]{Jun Yang}
\affiliation{Laboratory for Climate and Ocean-Atmosphere Studies, Department of Atmospheric and Oceanic Sciences, School of Physics, Peking University, Beijing 100871, China}
\affiliation{Institute of Ocean Research, Peking University, Beijing 100871, China}
\email{junyang@pku.edu.cn}  

\author[0000-0003-2278-6932]{Xianyu Tan}
\affiliation{Tsung-Dao Lee Institute, Shanghai Jiao Tong University, 1 Lisuo Road, Shanghai 200127, China}
\affiliation{School of Physics and Astronomy, Shanghai Jiao Tong University, 800 Dongchuan Road, Shanghai 200240, China}
\affiliation{State Key Laboratory of Dark Matter Physics, Shanghai Jiao Tong University, 1 Lisuo Road, Shanghai 200127, China}
\email{xianyut@sjtu.edu.cn}

\begin{abstract}

Tidally locked lava planets are hot, rocky worlds on close-in orbits with a permanent molten dayside. With JWST, their surfaces and atmospheres are beginning to be revealed. 
This work investigates 3D magma-ocean dynamics, derives scaling laws for the resulting ocean heat transport (OHT), and predicts its detectability. For the first time, the ocean circulation driven by the intense momentum and mass exchanges with the supersonic atmosphere is considered in addition to that by thermal forcing. The wind forcing turns out to overwhelmingly dominate the other two mechanisms, driving ocean currents reaching $\sim$100 m\,s$^{-1}$ and greatly expanding the latitudinal extent of the Matsuno-Gill response. 
Despite these extreme flow speeds, scaling analysis and 3D simulations consistently demonstrate that magma-ocean circulation alone does not produce an observable hotspot offset. This inefficiency arises because basin geometry and circulation structure fundamentally constrain zonal heat redistribution, suppressing large-scale longitudinal transport even under vigorous flow.

\end{abstract}



\section{Introduction}\label{sec:intro}

Tidally locked lava planets are close-in rocky worlds with substellar temperatures exceeding the silicate melting point ($\sim$1700~K; \citep{monteux2016cooling}), leading to hemispheric magma oceans \citep{leger2011extreme,batalha2011kepler,rouan2011,bourrier201855,malavolta2018ultra,zieba2022,brinkman2023toi,hu2024}. 
To date, more than one hundred such planets have been found\footnote{\url{https://exoplanetarchive.ipac.caltech.edu}}, among which, 55 Cnc e \citep{demory2016map,mercier2022}, K2-141 b \citep{zieba2022}, and Kepler-10 b \citep{hu2015semi} have been characterized through thermal phase curve observations. The results so far suggest no detectable hotspot displacement on K2-141 b and Kepler-10 b, whereas 55 Cnc e sometimes exhibits a significant eastward hotspot offset \citep{demory2016map,angelo2017}, and sometimes a negligible westward offset \citep{mercier2022}. A persistent eastward hotspot shift is generally interpreted as evidence for efficient atmospheric heat transport associated with Gill–Matsuno–type equatorial wave dynamics, as expected for planets with thick atmospheres. The absence of such a shift on K2-141 b and Kepler-10 b therefore suggests that these planets likely lack substantial atmospheres. In contrast, the episodic hotspot offsets reported for 55 Cnc e, if robust, point to additional mechanisms beyond steady atmospheric circulation. One plausible mechanism is heat transport driven by magma ocean circulation. Specifically, we are interested in whether magma ocean circulation can generate sufficiently strong heat transport to measurably influence thermal phase curves.

\begin{figure*}
    \centering
    \includegraphics[width=0.99\textwidth]{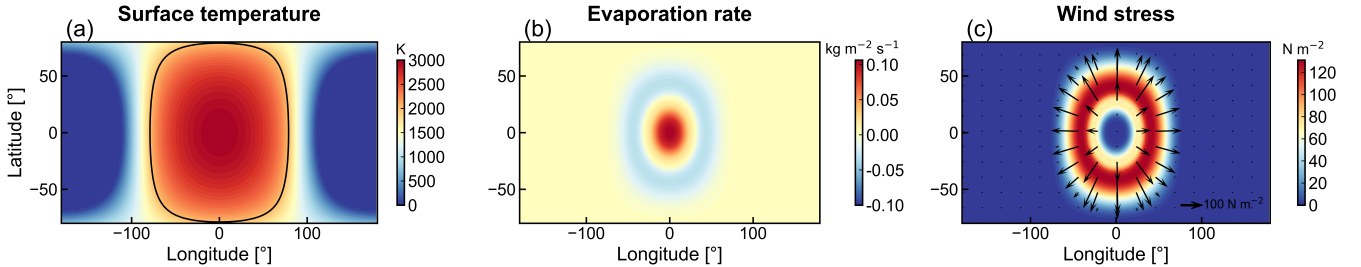}
    \caption{Drivers of ocean circulation on tidally locked lava worlds. (a) Surface temperature with a substellar temperature of 3000~K and a nightside temperature of 50~K. The ocean boundary, defined by the 2000~K liquidus, is indicated by the black contour. (b) Evaporation and condensation rates, with maxima of 0.1 and $-$0.05~kg\,m$^{-2}$\,s$^{-1}$, respectively; positive values correspond to evaporation (mass loss) and negative values to condensation (mass gain). (c) Horizontal wind stress, directed radially outward from the substellar point, peaking at approximately 120~N\,m$^{-2}$ near $\pm$40$^\circ$ in both longitude and latitude.}
    \label{fig:3forcing}
\end{figure*}

Magma oceans on tidally locked lava planets without a substantial atmosphere are simultaneously forced by three external drivers: thermal, wind, and evaporative forcing (Fig.~\ref{fig:3forcing}) \citep{leger2011extreme,castan2011atmospheres,nguyen2020modelling,kang2021escaping,nguyen2022impact}. Thermal forcing arises from the decay of surface temperature away from the substellar point (SP), sustaining horizontal pressure gradients in the ocean. In addition, the hot dayside maintains a significantly higher vapor pressure than the cold nightside, driving a supersonic atmospheric flow from day to night  \citep{castan2011atmospheres,nguyen2020modelling,kang2023true}.
This atmospheric flow imposes two additional forcings on the magma ocean: momentum drag at the ocean surface, which drives fluid away from the SP, and evaporation near the SP combined with deposition elsewhere, which generates a sea surface height (SSH) gradient that drives compensating interior circulation. Collectively, these forcings establish the large-scale circulation of the magma ocean, which in turn determines the surface temperature distribution.

\begin{table*}[tbp]
\caption{A summary of simulations. \label{tab:exps}}
\centering
\begin{tabular}{llll}
\hline\hline
 Group  & Substellar temperature $T_{\rm sub}$ &  Evaporation rate $E$ & Wind stress $\tau_s$ \\
\hline
All forcings & 3000 K & 0.1 kg\,m$^{-2}$\,s$^{-1}$ & 120 N\,m$^{-2}$ \\
Thermal-driven & [2500, 3000, 3500, 4000] K & $\times$0 & $\times$0\\
Evaporation-driven & 3000 K & $\times$ [0.5, 1, 2, 4] & $\times$0 \\
Wind-driven & 3000 K & $\times$0 & $\times$ [0.001, 0.01, 0.05, 0.1, 0.5, 1, 2] \\
\hline
\end{tabular}
\end{table*}

To quantify the magnitude of magma ocean circulation, an idealized two-dimensional (2D) non-rotating model was developed to investigate thermal- and wind-driven circulation as well as ocean depth \citep{lai2024a}. They found that the current speed can reach 0.1~m\,s$^{-1}$ (1.0~m\,s$^{-1}$) at the surface and 0.01~m\,s$^{-1}$ (0.1~m\,s$^{-1}$) in the interior under thermal (wind) forcing. Because these velocities are much smaller than the characteristic speed associated with planetary rotation ($\sim10^3$~m\,s$^{-1}$), rotation is expected to strongly modify the ocean dynamics. This motivates follow-up numerical studies that incorporate three-dimensional (3D) Coriolis effects \citep{yang2025}. Since the Coriolis force does no work, the magnitude of thermally driven flow in 3D remains comparable to that in 2D, but the flow pattern changes substantially, transitioning from a 2D overturning circulation to a Gill–Matsuno–type response, analogous to the atmospheric response \citep{showman2010}. 

Here we investigate 3D magma-ocean circulation on tidally locked lava planets using numerical simulations and scaling analysis. We characterize the circulation patterns produced by thermal, evaporative, and wind forcings and derive scaling laws for ocean current speed, heat transport, and ocean depth. These results provide a framework for predicting magma-ocean circulation and assessing the detectability of lava OHT across a wide range of planetary conditions.


\section{Method}\label{sec:method}
We simulate the 3D magma ocean circulation by modifying the MITgcm to solve the primitive equations in spherical coordinates with a free surface \citep{adcroft2004}. A primary technical challenge is that the depth and basin shape of the magma ocean are not known a priori. To address this, an offline coupling approach was employed \citep{yang2025}: 3D ocean simulations are first conducted under an arbitrary initial basin, and after reaching equilibrium, the basin shape is updated based on the simulated temperature. After $\mathcal{O}(10)$ such iterations, the solution converges.
Here, we address this challenge by disabling dynamics and advection through strong linear damping applied to temperatures below the liquidus (2000~K; \citep{monteux2016cooling}). This approach allows the basin boundary to evolve simultaneously with the temperature profile, producing robust results with minimal manual intervention. 

Our control planetary configuration is a planet of 2.8 Earth masses with a substellar temperature $T_{\rm sub} = 3000$ K. Fig.~\ref{fig:3forcing} shows the three external forcings imposed on the lava ocean. 
\begin{itemize}
\item \textbf{Thermal forcing.} Surface temperature is relaxed toward radiative equilibrium temperature over a radiative timescale of 4000~s (Eq.~\ref{eq:tau_rad}). The equilibrium temperature decays away from the SP following $T_{\rm sub}\cos^{1/4}\lambda\cos^{1/4}\phi$, where $\lambda$ and $\phi$ are longitude and latitude, with a minimum temperature of 50~K beyond $\pm110^{\circ}$ \citep{leger2009transiting}.
\item \textbf{Evaporative forcing.} Surface mass fluxes are computed using a one-dimensional (1D) atmospheric transport model \citep{kang2021escaping}. For the control configuration, peak evaporation occurs at the SP reaching approximately 0.1~kg\,m$^{-2}$\,s$^{-1}$, while peak condensation occurs near $ 40^{\circ}$ from the SP with a magnitude of 0.05~kg\,m$^{-2}$\,s$^{-1}$.
\item \textbf{Wind stress.} The wind stress is computed as $\tau_s=\rho_a \omega_a V_a$, where $V_a$ and $\rho_a$ are the atmospheric wind speed and density, respectively, and $\omega_a$ is the momentum exchange coefficient between the atmosphere and the lava surface \citep{ingersoll1985supersonic}. The coefficient $\omega_a$ is determined by two velocity components: $V_d$ and $V_e$. The first arises from turbulent eddies induced by vertical wind shear within the atmospheric boundary layer and is approximated as $V_d \sim C_d V_a$, where $C_d = 0.01$ representing the drag coefficient under high wind speeds \citep{sterl2017drag}. The second component arises from momentum deposition as supersonic vapor condenses onto the ocean surface, given by $V_e = \frac{mE}{\rho_a}$, where $m$ is the silicate molecular mass and $E$ is atmospheric condensation rate. $V_a$, $\rho_a$, and $E$ are computed using the 1D atmospheric transport model \citep{kang2021escaping}. For the control configuration assuming a SiO atmosphere, the peak atmospheric flow speed and condensation rate reach $\sim$3600~m\,s$^{-1}$ and $-0.05$ ~kg\,m$^{-2}$\,s$^{-1}$, respectively. This corresponds to a peak wind stress of roughly 120~N\,m$^{-2}$ near $\pm40^{\circ}$ in longitude and latitude, which is approximately three orders of magnitude larger than that over Earth’s oceans \citep{pedlosky1986buoyancy}.
\end{itemize}

We first perform simulations for the control planetary configuration (Table~\ref{si-table:paras}) with all forcings active, and then isolate the individual contributions of thermal, evaporative, and wind forcing. This approach allows us to assess the relative importance of each forcing type. To investigate how circulation strength, ocean depth, and heat transport respond to forcing amplitude, we additionally conduct experiments with different forcing magnitudes. Table~\ref{tab:exps} summarizes all experiments performed in this study. A more detailed description of the model setup, including the governing equations, parameters, and numerical setup, is provided in Appendix~\ref{si-sec:model}.
Typically, these simulations reach equilibrium after approximately 1000 years (Fig.~\ref{si-fig:initial_condition}). All results presented are 1000-day averages after the system attains quasi-equilibrium.

\begin{figure*}
    \centering
    \includegraphics[width=0.95\textwidth]{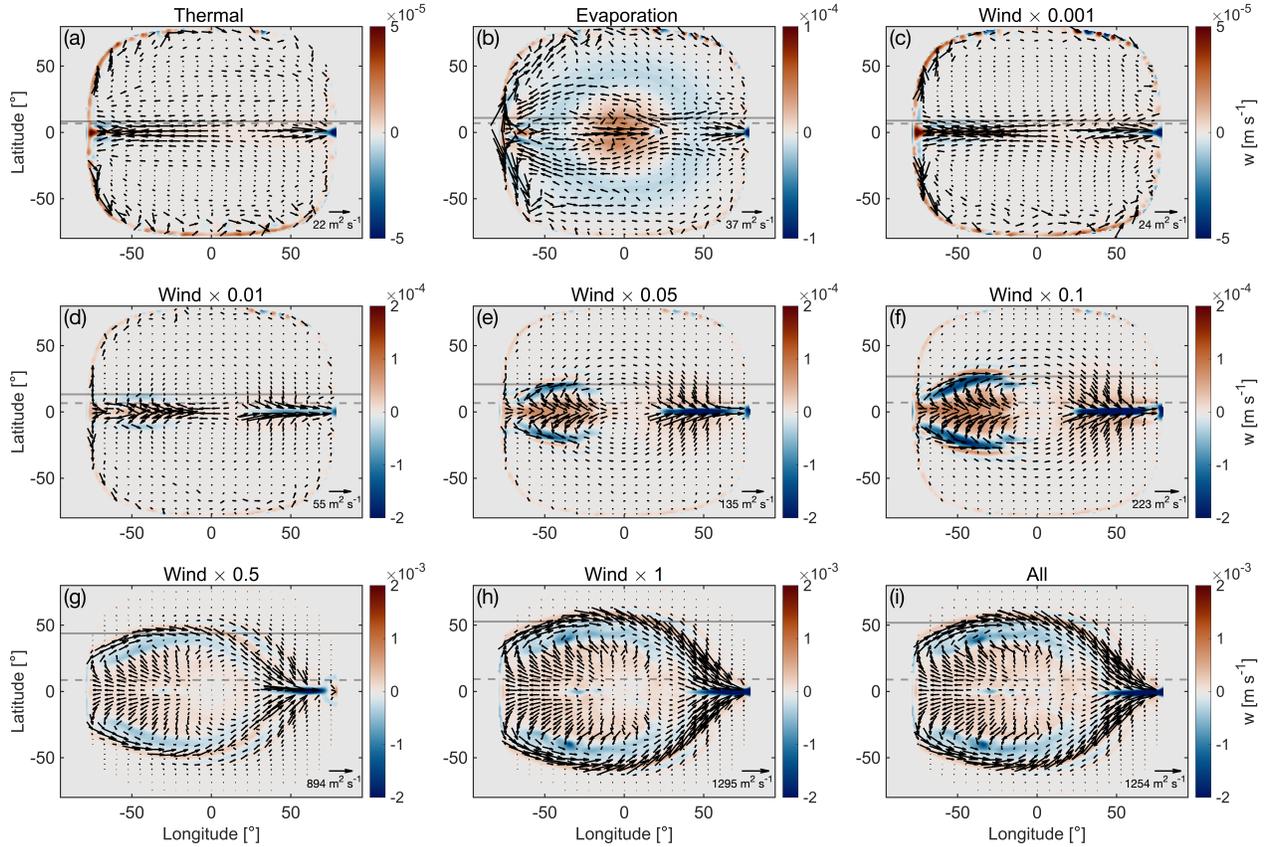}
    \caption{Ocean circulation driven by various forcings. Panels (a–i) show results driven by thermal forcing (a), evaporative forcing (b), wind forcing scaled by 0.001 (c), 0.01 (d), 0.05 (e), 0.1 (f), 0.5 (g), and 1 (h), and all forcings (i). Arrows represent vertically integrated horizontal flow over the upper 10 m (m$^2$\,s$^{-1}$), and colors represent vertical velocity at the base of the upper 10 m layer (m\,s$^{-1}$). 
    In each panel, the meridional extents set by the equatorial Rossby deformation radius, $L_{\beta}  = \sqrt{\sqrt{\Delta b D_o}/{\beta}}$, and the Rhines scale, $L_{\rm Rhines} = \sqrt{U/\beta}$, are indicated by gray dashed and solid lines, respectively. Here $\Delta b$ is the horizontal buoyancy contrast, $\beta$ is the meridional derivative of the Coriolis parameter, $D_o$ and $U$ are the characteristic magma ocean depth and horizontal current speed in the upper 10 m, respectively.
    A reference vector indicating the arrow length is shown in the bottom right corner of each panel.
    }
    \label{fig:3dvelocity}
\end{figure*}

\begin{figure*}
    \centering
    \includegraphics[width=0.95\textwidth]{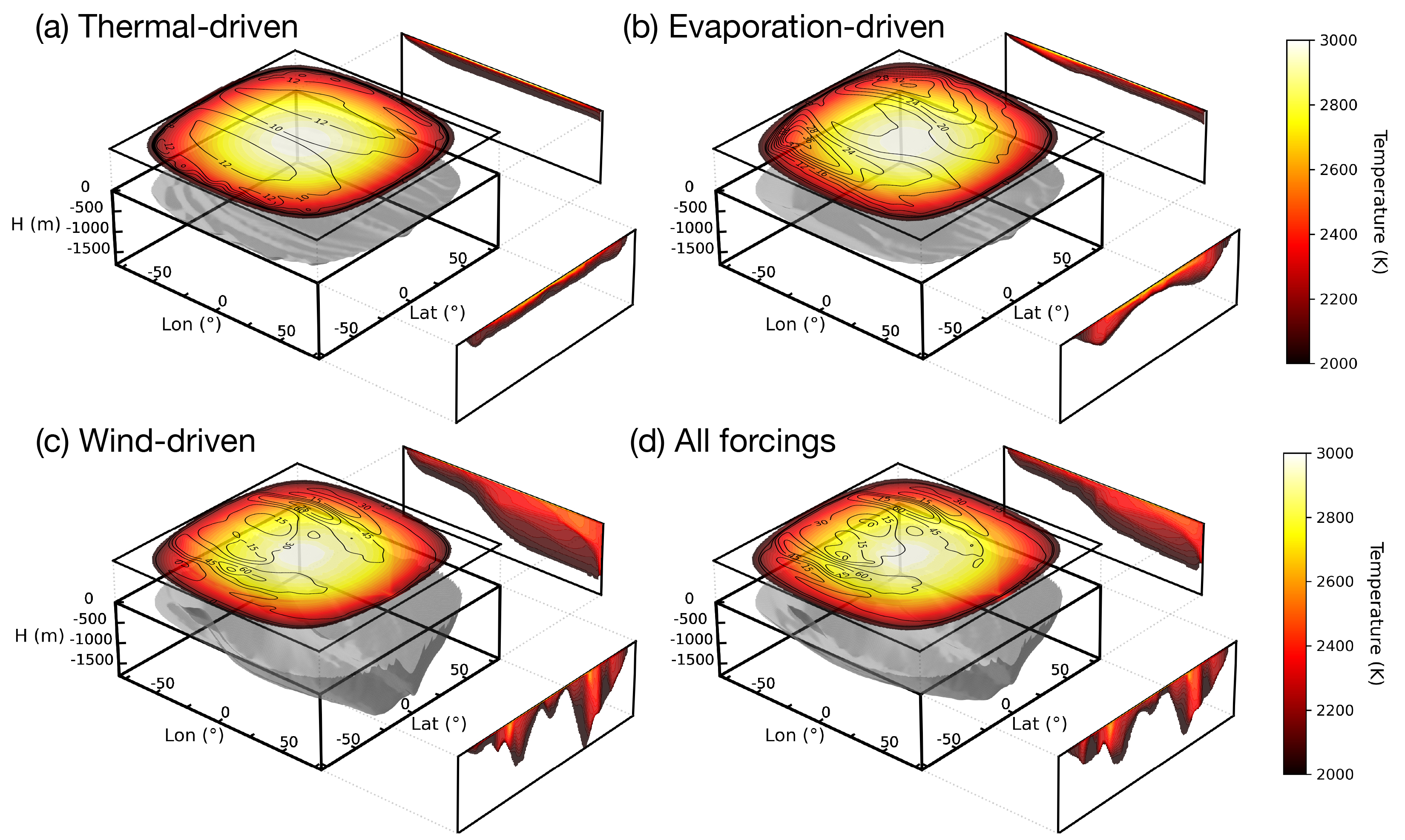}
    \caption{3D temperature structure and basin shape, and SSH driven by thermal (a), evaporative (b), wind (c), and all forcings (d). In each panel, colors indicate temperature, contours represent SSH (contour intervals of 1 m in (a), 4 m in (b), and 15 m in (c) and (d)), and gray shading masks the magma ocean basin. The surface temperature and SSH fields are projected onto the top plane as functions of longitude and latitude. A zonal–vertical temperature slice along the equator is projected onto the back plane, and a meridional–vertical slice along 50$^{\circ}$W is projected onto the right plane. For clarity, only the upper 1800~m is shown.
    }
    \label{fig:3dtemp}
\end{figure*}

\section{Results}\label{sec:result}

\subsection{Ocean circulation pattern and basin bathymetry}\label{sec:circ}

We first present the circulation pattern (Fig.~\ref{fig:3dvelocity}) and discuss the underlying physics. 
Regardless of the type of external forcing, Gill response is expected in the equatorial regions \citep{matsuno1966,gill1980}. According to linear theory, the meridional extent of the Gill response is set by the equatorial Rossby deformation radius \citep{showman2011}, 
\begin{equation}
    L_{\beta}  = \sqrt{\frac{\sqrt{\Delta b D_o}}{\beta}}\sim 2\times10^6 ~ \rm  m, 
\end{equation}
where $\Delta b$ is the horizontal buoyancy contrast, $\beta$ is the meridional derivative of the Coriolis parameter, and $D_o$ is the characteristic magma ocean depth. For the control configuration, $L_{\beta} $ is around 2000~km (10$^\circ$ in latitude), as indicated by the gray dashed lines in Fig.~\ref{fig:3dvelocity}. Consistent with this scaling, the thermal- and evaporation-driven circulations indeed present a very narrow equatorial circulation constrained within $\pm10^{\circ}$ latitude (Fig.~\ref{fig:3dvelocity}(a-b)).

In contrast, the Gill response in the wind-driven case is more pronounced in both strength and latitudinal extent (Fig.~\ref{fig:3dvelocity}(e–h)). This difference arises not from the spatial structure of the wind forcing—which, like thermal forcing, drives flow away from the SP—but from its magnitude. To test this, we reduce the wind forcing by a factor of 1000, producing a circulation (Fig.~\ref{fig:3dvelocity}(c)) that closely resembles the thermally driven case (Fig.~\ref{fig:3dvelocity}(a)). We then gradually increase the wind-forcing amplitude and find that the latitudinal extent of the Gill response expands accordingly (Fig.~\ref{fig:3dvelocity}(c–h)).

The oversized Gill response stems from nonlinear effect. 
Since wind forcing is substantially stronger than thermal and evaporative forcings, the circulation strengthens accordingly, increasing nonlinearity and altering the meridional extent of the equatorial circulation. In this regime, the meridional width of the Gill response is set by the Rhines scale \citep{Rhines_1975},
\begin{equation}
    L_{\rm Rhines}  = \sqrt{\frac{U}{\beta}}, 
\end{equation}
where $U$ is the characteristic flow speed. As wind forcing magnitude increases from 0.001$\times$ to 1$\times$ the default value, $L_{\rm Rhines}$ varies from $2\times10^6$ to $10^7$~m, as shown in the gray solid lines in Fig.~\ref{fig:3dvelocity}. Since $L_{\rm Rhines} \gg L_{\beta}$ in the experiments with wind forcing $\ge 0.05$ times the reference value, the widths of the equatorial circulation scale accordingly with $L_{\rm Rhines}$, as evident in Fig.~\ref{fig:3dvelocity}.

Beyond the tropical region, set by $\max{(L_{\beta}, L_{\rm Rhines}})$, geostrophic flow forms in the ocean interior, and strong return currents appear along the western boundary in the thermal- and evaporative-forced experiments. A similar circulation exists in Earth’s oceans at mid-latitudes: anticyclonic atmospheric wind stress drives an equatorward interior flow by consuming the ocean’s potential vorticity, while the return flow becomes concentrated along the western boundary, where frictional drag restores potential vorticity \citep{stommel1948,munk1950,vallis2017atmospheric}.
Similarly, in the lava ocean, mass redistribution and thermal forcing modify potential vorticity by directly inducing a vertical velocity near the surface, $w_f$. This vertical velocity then drives a meridional flow $v$ through the Sverdrup relationship \citep{vallis2017atmospheric}:
\begin{equation}
\begin{aligned}
\beta v &= f \frac{\partial w}{\partial z}
          \sim \frac{f w_f}{D_o}, \\
    w_f &\sim \left( \frac{k_z^2 \Delta b}{f L^2} \right)^{1/3} \; \text{(thermal)} \;, ~\sim \frac{E}{\rho_0} \; \text{(evaporative)}.
\end{aligned}
\label{eq:sverdrup}
\end{equation}
Here, $f$ is the Coriolis parameter, $\rho_0$ is the reference ocean density, $k_z$ is the vertical diffusivity, and $L$ is the characteristic horizontal scale of the magma ocean.
Substituting the parameters adopted in our experiments (Table~\ref{si-table:paras}), we estimate a meridional flow speed of $v \sim 0.01$~m\,s$^{-1}$ for the thermally forced experiment and $v \sim 1.0$~m\,s$^{-1}$ for the evaporation-forced experiment. These estimates are broadly consistent with the meridional flow speeds obtained in our numerical simulations (Fig.~\ref{fig:3dvelocity}(a–b)).

It is also worth noting that the extratropical circulation patterns differ substantially between the thermally forced and evaporative-forced experiments. This difference arises from their distinct $w_f$ distributions (colors in Fig.~\ref{fig:3dvelocity}(a-b)). Thermal forcing increases the column height near the SP through heat deposition; at equilibrium, dynamic column compression ($\partial_zw<0$) is required, inducing an equatorward interior flow ($v<0$) and a poleward boundary current. In contrast, evaporative forcing removes mass near the SP and deposits it in a band roughly $40^\circ$ away (Fig.~\ref{fig:3forcing}(b)), driving a poleward interior flow near the SP within $40^\circ$ (Eq.~\ref{eq:sverdrup} and Fig.~\ref{fig:3dvelocity}(b)). Beyond this region, the circulation reverses as the flow transitions from the evaporative zone to the mass-deposition zone.

Across all experiments, regardless of the type of forcing, divergent flow prevails in the equatorial regions as part of the Gill response. This drives an upwelling motion near the equator and a downwelling return flow near the eastern coast (shading in Fig.~\ref{fig:3dvelocity}). The upwelling draws cold fluid upward from the bottom, reducing the ocean depth near the equator (Fig.~\ref{fig:3dtemp}). The downwelling near the eastern coast subsequently bifurcates into two branches, one in each hemisphere, as it sinks. By advecting surface heating downward, this flow sculpts two deep channels flanking the equatorial shoal.

To conclude this section, we note that the 3D circulation obtained here differs markedly from what is reported by \cite{lai2024a,lai2024b}, where a 2D framework is adopted. While the 2D scaling still provides useful insight into the energetics, since the Coriolis force does no work, the circulation pattern and basin bathymetry produced in a 2D framework miss key dynamic features, including the Gill pattern near the equator and the boundary currents in the mid-latitudes. As will be shown in the following section, planetary rotation and the $\beta$-effect (i.e., the variation of the Coriolis force with latitude) play a central role in setting the ocean depth.
Our results also differ from the 3D numerical study of \cite{yang2025}, where only thermal forcing is considered. For a SiO atmosphere, atmospheric wind speed can reach $\mathcal{O}(1000)$~m\,s$^{-1}$, inducing a wind stress of $\sim100$~N\,m$^{-2}$, approximately three orders of magnitude larger than that on Earth \citep{pedlosky1986buoyancy}. This wind stress dominates the forcing budget and becomes the primary driver of lava-ocean dynamics (Fig.~\ref{fig:3dvelocity}(i) and Fig.~\ref{fig:3dtemp}(d)).

\begin{table*}[tbp]
\caption{Summary of Scaling Laws under Varying Forcings}
\label{tab:scaling}
\centering
\setlength{\tabcolsep}{4pt} 
\resizebox{0.9\linewidth}{!}{
\begin{tabular}{lcccc}
\hline
Forcing         &  $U$                & $D$                  &   $\Delta \eta$                  &  $H$ \\ \hline
Thermal       & $(\frac{k_z \Delta b^2}{f^2 L})^{1/3}$ & $(\frac{ fk_z L^2}{\Delta b})^{1/3}$  & $(\frac{fk_zL^2 \Delta b^2}{g^3})^{1/3}$ & $\frac{\rho c_p}{g\alpha} (\frac{k_z^2 \Delta b^4 }{f L^2})^{1/3}$ \\ 
Evaporative & $(\frac{E g}{\rho f})^{1/2}$     & $(\frac{EfL^2 g}{\rho \Delta b^2})^{1/2}$ & $(\frac{EfL^2}{\rho g})^{1/2}$       & $\frac{c_p E}{\alpha}$  \\
Wind              & $(\frac{\tau_s \Delta b}{\rho f^2 L})^{1/2}$  & $(\frac{\tau_s L}{\rho \Delta b})^{1/2}$    & --                        & $\frac{c_p \tau_s \Delta T}{fL}$  \\ 
\cline{1-5}              
\end{tabular}
}
\vspace{1mm}
\footnotesize
\\
\parbox{\linewidth}{%
\textbf{Note 1.} $U$, $\Delta \eta$, $D$, and $H$ are the characteristic scales of horizontal velocity, SSH difference, thermocline depth, and OHT divergence. The scaling for ocean depth $D_o$ is not shown, as it is similar to the thermocline depth $D$ (see Eq.~\ref{eq:Do}). In a wind-driven regime, no scaling law is provided for $\Delta \eta$ because the surface velocity is driven by wind stress rather than by a pressure gradient.\\
\textbf{Note 2.} $\Delta b = g\alpha\Delta T$ is horizontal buoyancy contrast, where $g$ is gravity, $\alpha$ is thermal expansion coefficient, $\Delta T = T_{\rm sub} - T_{\rm liq}$, $T_{\rm sub}$ and $T_{\rm liq}$ are substellar temperature and the liquidus. $\tau_s\sim C_d \rho_a V_a^2 \propto P_a$ is surface wind stress (Eq.~\ref{eq:stress-scaling}) and $E \sim P_a V_a/(ga) \propto P_a/a^2$ is evaporation rate (Eq.~\ref{eq:evaporation-scaling}), where $P_a$ is the saturated vapor pressure, $V_a$ and $\rho_a$ are atmospheric wind speed and density, $a$ is planetary radius, and $C_d$ is surface drag coefficient. 
$L = a\theta_p$ is the horizontal scale of the magma ocean, where $\theta_p \approx \cos^{-1} ( \frac{T_{\rm liq}}{T_{\rm sub}})^4$ is the angular radius of the magma ocean. $f$ is the Coriolis parameter, $\beta$ is the meridional derivative of the Coriolis parameter, $k_z$ is vertical diffusivity, $\rho$ is fluid density, and $c_p$ is heat capacity at constant pressure. The default values of planetary and oceanic parameters used in the scaling laws are presented in Table~\ref{si-table:paras}.
}
\end{table*}

\begin{figure*}
    \centering
    \includegraphics[width=0.98\textwidth]{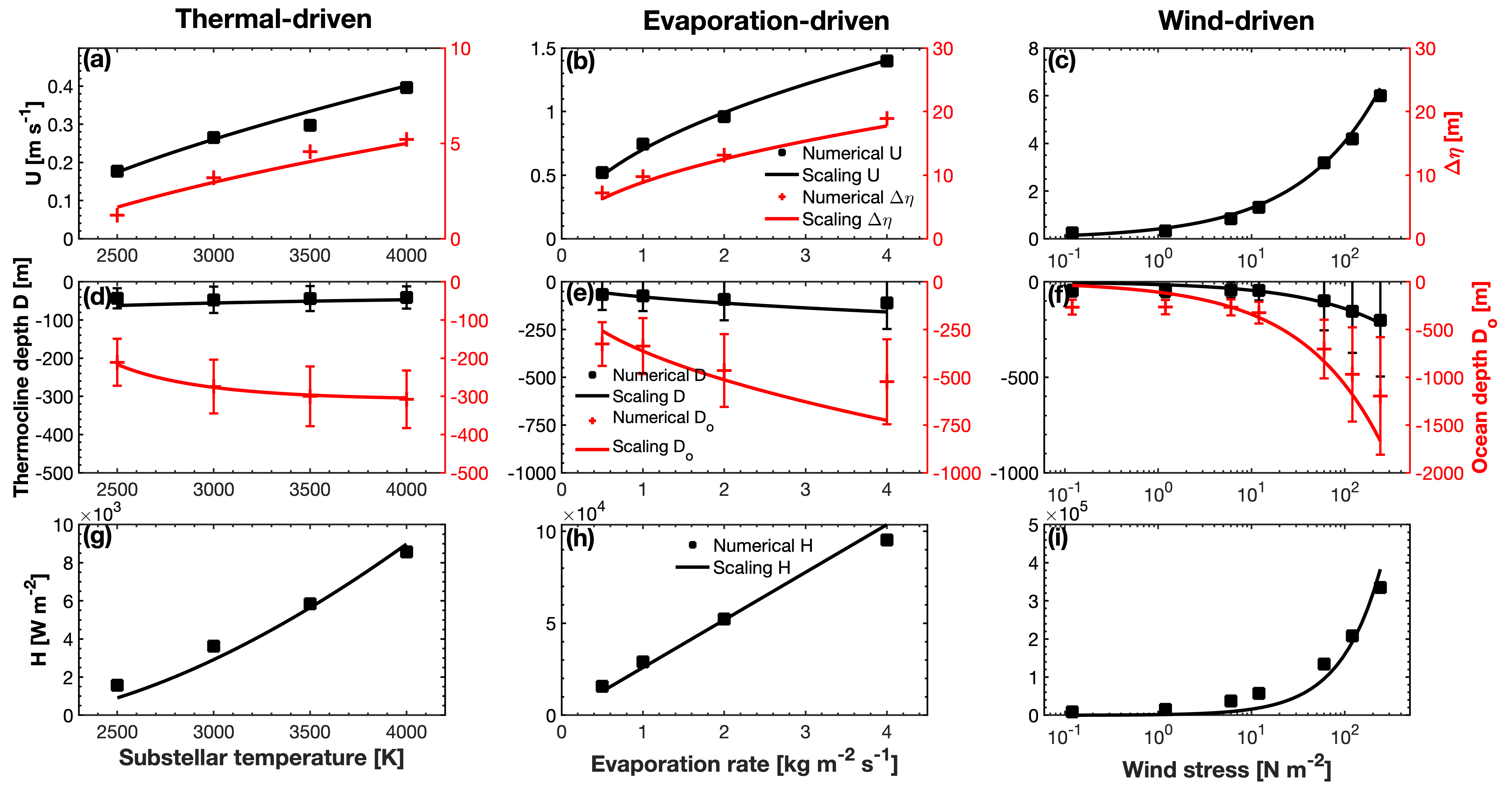}
    \caption{Examination of scaling laws under thermal-driven (left), evaporation-driven (middle), and wind-driven (right) conditions. From top to bottom, panels show horizontal velocity ($U$, black squares and lines) and SSH difference ($\Delta \eta$, red crosses and lines); thermocline depth ($D$, black squares and lines) and magma ocean depth ($D_o$, red crosses and lines), with the corresponding $\pm1\sigma$ intervals indicated by error bars; and OHT divergence ($H$) as functions of forcing amplitude. Under wind-driven conditions, no scaling law is shown for $\Delta \eta$ because the surface velocity is forced by wind stress rather than by a pressure gradient. 
    In all panels, markers represent mean values within the magma ocean obtained from the 3D simulations, while solid lines indicate predictions from scaling laws. }
    \label{fig:scaling}
\end{figure*}

\subsection{Scaling analysis}\label{sec:scaling}

Building upon the qualitative physical understanding from the previous section, we here derive quantitative scaling laws for the horizontal velocity ($U$), SSH difference ($\Delta \eta$), thermocline depth ($D$), magma ocean depth ($D_o$), and ocean heat transport divergence ($H$) under the three types of forcing. These scaling laws enable predictions for arbitrary lava-planet configurations without the need for additional numerical simulations.

Here, we consider the dominant balance in the momentum and heat budgets (Fig.~\ref{si-fig:mom-thermo}), together with mass continuity, from which we solve $U$, $\Delta \eta$, $D_o$/$D$, and $H$ as functions of planetary radius $a$, gravity $g$, rotation rate $\Omega$ (i.e., the Coriolis parameter $f$), vertical diffusivity $k_z$, SP temperature $T_{\rm sub}$, wind stress $\tau_s$, and evaporation rate $E$.

When thermal forcing dominates, the scaling is based on leading-order balances among the following four equations. 
\begin{enumerate}
    \item \textbf{Mass continuity:} This relates horizontal and vertical velocities through the thermocline depth:
    \begin{equation}
        \frac{U}{L} \sim \frac{W}{D},
        \label{eq:mass}
    \end{equation}
    where $L = a \theta_p$ is the horizontal scale of the magma ocean, $\theta_p = \cos^{-1} \left[\left(\frac{T_{\rm liq}}{T_{\rm sub}}\right)^4\right]$ is the angular radius of the magma ocean, determined by the substellar temperature $T_{\rm sub}$ and the liquidus $T_{\rm liq}$.

    \item \textbf{Geostrophic balance in the interior:} The Coriolis force balances the pressure gradient force associated with vertically integrated horizontal buoyancy contrast:
    \begin{equation}
        f U \sim \frac{\Delta b D}{L},
        \label{eq:geos}
    \end{equation}
    where $\Delta b = g \alpha \Delta T$ is the horizontal buoyancy contrast, with $\alpha$ the thermal expansion coefficient, and $\Delta T = T_{\rm sub} - T_{\rm liq}$ the horizontal temperature contrast.

    \item \textbf{Surface geostrophic balance:} In the absence of wind forcing, the surface velocity is driven by the SSH gradient:
    \begin{equation}
        f U \sim \frac{g \Delta \eta}{L}.
        \label{eq:geos_surf}
    \end{equation}
    Following geostrophic balance, ocean currents tend to flow along isobars, forming closed gyres or jets depending on the spatial structure of $\Delta \eta$ (Fig.~\ref{fig:3dvelocity}(a-b)).

    \item \textbf{Advection-diffusion balance:} In thermally-forced cases, dense surface fluid near the ocean boundary sinks and accumulates at the ocean bottom, while vertical diffusion returns it upward. At equilibrium, the downward diffusive heat flux is balanced by upward advective transport in the upwelling regions \citep{vallis2019essentials}:
    \begin{equation}
        D \sim \frac{k_z}{W},
        \label{eq:adv-diff}
    \end{equation}
    where $W$ is the characteristic scale of vertical velocity.
\end{enumerate}
Together, these four leading-order balances yield the thermal-driven scaling laws. 

In the evaporation-driven cases, mass sources and sinks shape the SSH gradient. Evaporation reduces SSH and condensation increases SSH, while horizontal advection removes this anomaly. At equilibrium:
\begin{equation}
    U\frac{\Delta \eta}{L} \sim \frac{E}{\rho}, 
    \label{eq:mass_ssh}
\end{equation}
where $\rho$ is fluid density. Combining Eqs.~\ref{eq:mass}-\ref{eq:geos_surf} and Eq.~\ref{eq:mass_ssh} gives the evaporation-driven scaling laws.

When wind forcing dominates, the system still follows mass continuity (Eq.~\ref{eq:mass}) and interior geostrophic balance (Eq.~\ref{eq:geos}). Besides, the vertically-integrated momentum equation gives
\begin{equation}
    \rho f U D \sim \tau_s,
    \label{eq:surfacemom_wind}
\end{equation}
where $\tau_s$ is wind stress. The scaling laws for the wind-driven regime are obtained by combining Eqs.~\ref{eq:mass}-\ref{eq:geos} and Eq.~\ref{eq:surfacemom_wind}. Notably, no scaling law is derived for $\Delta \eta$ in this regime because surface flow is directly forced by wind stress, rather than SSH gradient.
The scaling laws under different dominant forcings are summarized in Table \ref{tab:scaling}. 

Assuming temperature decreases with depth exponentially within the magma ocean, with surface temperature of $T_{\rm sub}$ and bottom temperature of $T_{\rm liq}$, the vertical temperature profile can be given as
\begin{equation}
T = T_{\rm liq} + (T_{\rm sub} - T_{\rm liq}) \cdot \rm exp({-z/D}),
\label{eq:Tz}
\end{equation}
where $z$ denotes depth measured downward from the surface (positive), and $D$ is the thermocline depth, defined as the e-folding depth at which the temperature contrast has decayed to $1/e$ of the surface-bottom contrast, i.e., $T_{\rm sub} - T(z = -D) = (T_{\rm sub} - T_{\rm liq})/e$. The ocean depth $D_o$ is defined as the depth where the temperature reaches a fixed threshold $T_o$=2100~K, slightly above the liquidus. Accordingly, the ocean depth can be approximated as the thermocline depth multiplied by a coefficient that depends on the substellar temperature:
\begin{equation}
D_o \sim D \cdot ln(\frac{T_{sub}-T_{liq}}{T_o-T_{liq}}).
\label{eq:Do}
\end{equation}
Finally, the OHT divergence can be estimated as 
\begin{equation}
\label{eq:oht}
H \sim \rho c_p \Delta T \frac{UD_o}{L},
\end{equation}
where $c_p$ is the heat capacity at constant pressure. 

Fig.~\ref{fig:scaling} compares the scaling laws with the numerical simulations and shows overall good agreement. Regardless of the forcing type, increasing the forcing amplitude leads to faster horizontal flow ($U$), larger SSH perturbations ($\Delta \eta$), a deeper magma ocean ($D_o$), and enhanced OHT divergence ($H$), consistent with physical intuition. One exception is the thermocline depth $D$, which decreases with increasing thermal forcing amplitude. This behavior arises because, although thermal forcing energizes the circulation, it simultaneously suppresses vertical motions by establishing strong vertical stratification. Despite the shallower thermocline, the total magma ocean depth $D_o$ still increases with $T_{\rm sub}$, because higher surface temperatures render the 2000~K liquidus effectively colder, pushing the liquidus interface to greater depths (Eq.~\ref{eq:Do}).
Comparing the magnitude of $U$ driven by different forcings, we notice that the wind-driven circulation dominates over both thermal- and evaporation-driven flows. This is consistent with the simulation results that, when all three types of forcings are applied simultaneously, the fluid pattern resembles the wind-driven case (Fig.~\ref{fig:3dvelocity}(i)). 

\begin{figure*}
    \centering
    \includegraphics[width=0.98\textwidth]{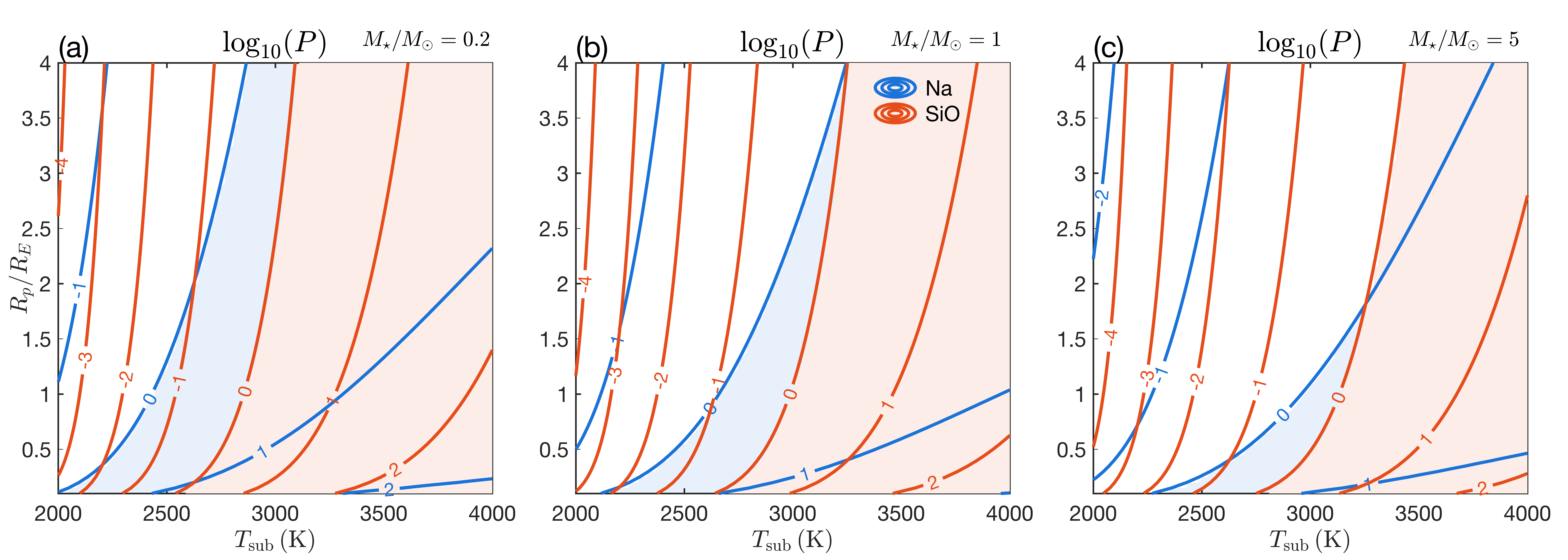}
    \caption{Detectability index, $\log(\mathcal{P})$, under a range of stellar, planetary, and atmospheric conditions. From left to right, panels show $\log(\mathcal{P})$ as a function of planetary radius $R_p/R_E$ and substellar temperature $T_{\rm sub}$ for stellar mass ratios of $M_{\star}/M_{\odot} = 0.2$, 1, and 5, respectively. Here, $R_p$ and $R_E$ denote the radii of the lava planet and Earth, $M_\star$ and $M_{\odot}$ denote the stellar and solar mass. In each panel, blue and orange contours orrespond to Na and SiO vapor atmospheres, respectively, and the shaded regions indicate regimes in which lava OHT is detectable, defined by $\log(\mathcal{P})>0$. }
    \label{fig:detect_index}
\end{figure*}

Across all cases examined here, the magma-ocean depth $D_o$ remains below 2000~m. Limited by this shallow depth, the mean OHT divergence $H$ within the ocean ranges from $10^{3}$ to $4\times10^5$ W\,m$^{-2}$. Even the highest OHT considered here is small compared to Kepler-10 b’s insolation ($\sim 5 \times 10^6$ W\,m$^{-2}$). As a result, the surface temperature remains close to the prescribed radiative equilibrium (colors in Fig.~\ref{fig:3dtemp}), leading to only a very weak signal in thermal phase curve (see Fig.~\ref{si-fig:fpfs}).

\section{Discussion}\label{sec:discuss}
Kepler-10 b’s phase curve is unlikely to be significantly modified by OHT (Fig.~\ref{si-fig:fpfs}); however, this does not imply the same for other lava planets. Using the scaling laws for OHT divergence (Table \ref{tab:scaling}), we can predict the strength of $H$ for any lava planet given its planetary parameters. Here, we define a detectability index $\mathcal{P}$ as the ratio of the wind-driven OHT divergence $H$ to the stellar insolation $S_0$,
\begin{equation}
    \mathcal{P}\equiv \frac{H}{S_0}\sim \frac{c_p \tau_s \Delta T}{fL} \frac{1}{\sigma T_{\rm sub}^4},
    \label{eq:wind_effect}
\end{equation}
where $\sigma$ is the Stefan-Boltzmann constant.
Substituting the scaling relations for wind stress and atmospheric properties,  $\tau_s \sim \rho_a C_d V_a^2$, $V_a\sim \sqrt{\gamma RT_{\rm sub}}$ \citep{kang2023true}, and $\rho_a\sim P_a/(RT_{\rm sub})$, we get
\begin{equation}
    \mathcal{P}\sim \frac{\gamma C_d c_p}{\sigma} \frac{P_{\rm eq}(T_{\rm sub})}{fa T_{\rm sub}^3},
\end{equation}
Here, $L\sim a$ is planetary radius, $\gamma\approx1.4$ is ratio of specific heats, $R$ is specific gas constant, and $P_a$ is estimated by the equilibrium pressure $P_{\rm eq}$ at the substellar point. 

Further note that the rotation rate $\Omega\equiv f/2$, is related to $T_{\rm sub}$ through $\Omega = \sqrt{\frac{GM_{\star}}{d^3}}$ and $T_{\rm sub} = (\frac{L_{\star}}{\sigma d^2})^{1/4}$, where $L_{\star}=L_{\odot}(\frac{M_\star}{M_{\odot}})^4$ for main-sequence stars \citep{kang2023true}, $d$ is the semi-major axis, $M_{\star}$ and $L_{\star}$ are the stellar mass and luminosity, $M_{\odot}$ and $L_{\odot}$ are the mass and luminosity of the Sun, and $G$ is the gravitational constant. We get the following dependence of $\mathcal{P}$ on fundamental planetary parameters:
\begin{equation}
    \mathcal{P} \sim \gamma C_d c_p \frac{L_{\odot}^{3/4}}{\sigma^{7/4}G^{1/2}} \frac{P_{\rm eq}(\rm{chem}, T_{\rm sub})}{a M_{\star}^{1/2}T_{\rm sub}^6}, 
    \label{eq:wind_effect_planetary_para}
\end{equation}
where the equilibrium vapor pressure $P_{\rm eq}$ depends on the chemical composition and $T_{\rm sub}$ \citep{nguyen2020modelling}.
Eq.~\ref{eq:wind_effect_planetary_para} indicates that the magnitude of the temperature perturbation associated with wind-driven ocean circulation increases with the vapor pressure $P_{\rm eq}$, but decreases with increasing planetary radius $a$, stellar mass $M_{\star}$, and substellar temperature $T_{\rm sub}$.

\begin{figure}
    \centering
    \includegraphics[width=0.95\linewidth]{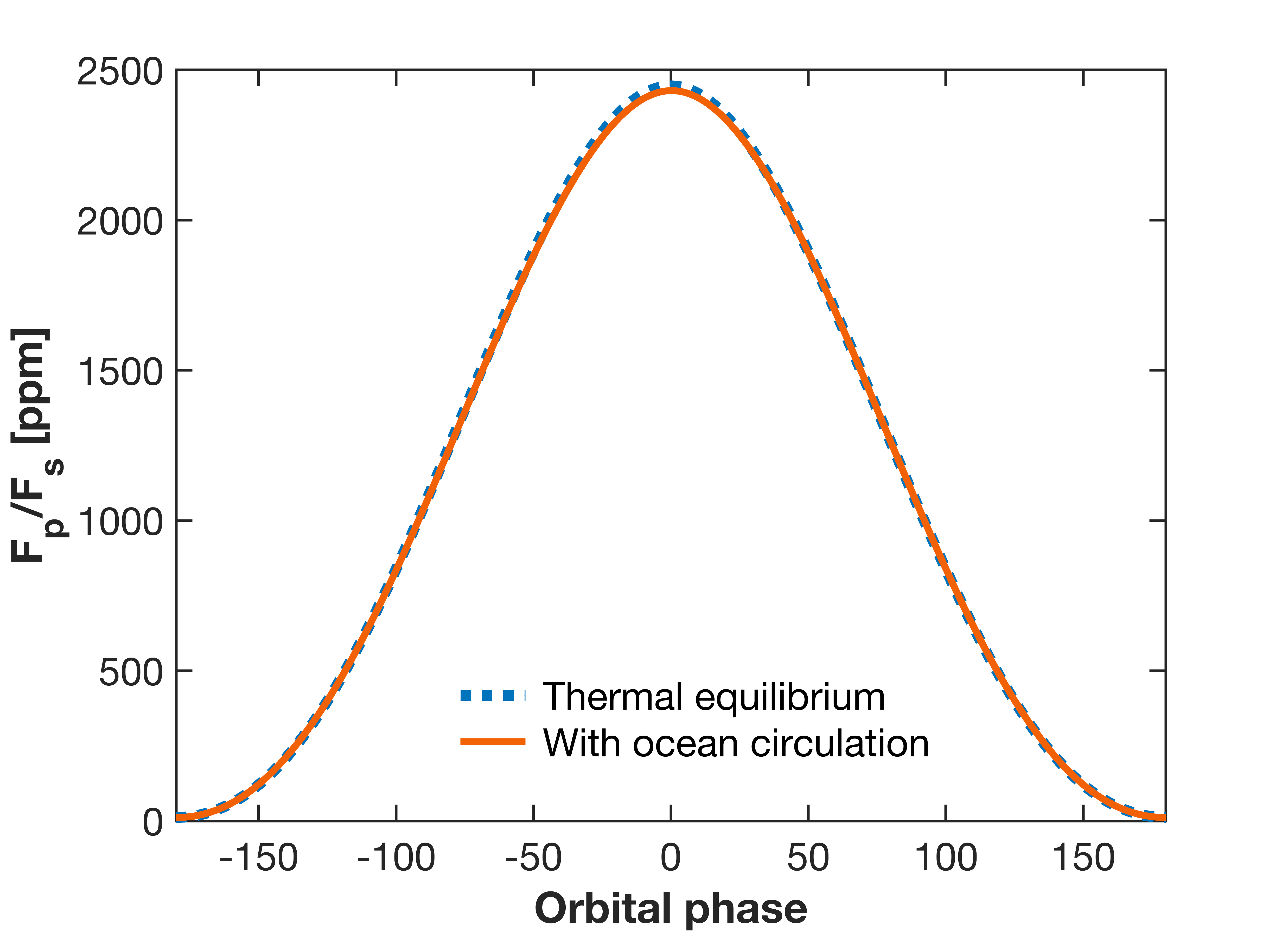}
    \caption{Thermal phase curves for a sodium atmosphere with $T_{\rm sub} = 3000$ K on a Earth-size planet orbitng an M-dwarf star. The dotted blue line represents radiative equilibrium (no circulation), and the solid orange line includes wind-driven circulation. The two curves nearly overlap, indicating that lava OHT does not produce a detectable phase offset.}
    \label{fig:fpfs}
\end{figure}

Fig.~\ref{fig:detect_index} shows $\log(\mathcal{P})$ as a function of planetary radius $R_p/R_E$ and substellar temperature $T_{\rm sub}$ for various stellar mass ratios, assuming sodium (blue) and SiO (orange) atmospheres. Here, $R_E$ and $M_{\odot}$ denote Earth's radius and the solar mass, respectively. Lava OHT is detectable in the shaded areas where $\log(\mathcal{P}) \ge 0$. 
For a given star mass, $\mathcal{P}$ generally increases with decreasing planetary radius/mass and with higher substellar temperatures, due to weaker ocean stratification on smaller planets and the rapid exponential rise of vapor equilibrium pressure with temperature \citep{nguyen2020modelling,kang2021escaping}. 
Sodium atmospheres tend to produce larger $\mathcal{P}$ values than SiO, particularly at relatively lower substellar temperatures, because of sodium’s higher volatility.
Comparing different stellar masses, lava OHT detectability is higher for planets orbiting less massive stars. This occurs because a lower stellar mass implies a smaller planetary rotation rate, which results in stronger ocean current speed (Eqs.~\ref{eq:geos} and \ref{eq:surfacemom_wind}) and consequently larger $\mathcal{P}$.

To test this scaling, we performed an additional simulation with a sodium atmosphere and $T_{\rm sub}=3000$ K on an Earth-size planet orbiting an M-dwarf ($M_\star = 0.2\,M_\odot$). This configuration lies within the detectable regime in Fig.~\ref{fig:detect_index}. Contrary to expectation, however, no measurable phase offset emerges. Although the planet–star flux ratio increases substantially, the phase curve remains nearly identical to that under radiative equilibrium (Fig.~\ref{fig:fpfs}).
This result demonstrates that strong OHT does not necessarily imply longitudinal heat redistribution. The scaling captures only the magnitude of OHT, whereas the emergence of a phase offset depends on the efficiency of zonal heat transport, which is controlled by the spatial structure of the circulation.

Previous studies have shown that significant phase offsets can arise when oceanic or atmospheric superrotation develops on tidally locked planets around M-dwarf stars, typically in the absence of lateral boundaries and under conditions favorable for large-scale wave-mean flow interactions \citep[e.g.,][]{showman2011,imamura2020,hu2014,zeng2021}. Such superrotation and longitudinal heat transport fail to develop in our model for the following three reasons.
First, the solid boundaries of the ocean basin inhibit the development of a deep, penetrating eastward jet, thereby suppressing efficient zonal heat transport. Second, the Matsuno-Gill response is confined by basin geometry. When the Rossby gyre expands beyond the size of the basin (i.e., $L_\beta\gtrsim L$), ocean currents are forced to follow the basin boundary, a nearly isothermal pathway, further limiting net heat redistribution. Third, the large atmospheric Rossby number ($R_o \sim 5$) suppresses the propagation of planetary-scale Rossby waves and the associated momentum convergence required to drive equatorial superrotation \citep{showman2011,imamura2020}. The absence of superrotation in the atmosphere circulation does not help drive the superrotation in the ocean, which in principle should have induced a substantial hotspot displacement \citep{zeng2021}.
More generally, capturing this redistribution efficiency within a predictive scaling framework would require resolving the nonlinear, boundary-constrained dynamics of the circulation, which lies beyond the scope of the present study and remains an open problem for future work.

Taken together, the simulations and scaling analyses demonstrate that lava ocean circulation driven by thermal, evaporative, and wind forcing alone cannot generate an observable hotspot shift on tidally locked lava planets. Consequently, the detection of a significant phase curve offset on a lava planet would instead point to the presence of a thick atmosphere, rather than a tenuous vapor envelope. This conclusion is consistent with current observations of 55 Cnc e \citep{demory2016map,hu2024}.


It is noteworthy that although our simulations consistently show that wind forcing dominates over thermal and evaporative forcing under the adopted planetary configuration, this hierarchy depends on atmospheric and planetary parameters. To generalize the dynamical regime, we express the wind stress $\tau_s$ and evaporative flux $E$ in terms of bulk planetary properties. The wind forcing scales as following 
\begin{equation}
  \label{eq:stress-scaling}
  \tau_s\sim C_d\rho_a V_a^2\sim C_d\gamma P_a\propto P_a.
\end{equation}
In turn, the evaporative forcing $E$ can be estimated by the total atmospheric transport divided by the surface area. Adopting the hydrostatic balance $\int_z \rho_a~dz=P_a/g$ and the fact that $g\propto a$ with the planet's bulk density fixed, we get
\begin{equation}
  \label{eq:evaporation-scaling}
  E\sim P_aV_a/(ga)\sim P_a\sqrt{\gamma RT_{\rm sub}}/(ga) \propto P_a/a^2.
\end{equation}
From Eqs.~\ref{eq:stress-scaling}-\ref{eq:evaporation-scaling} we know that lava oceans on planets with more refractory atmosphere will be dominantly driven by the thermal forcing, and lava oceans on small planets with volatile atmosphere will be dominantly driven by evaporative forcing.

\section{Summary}\label{sec:conclude}

We modified MITgcm to simulate 3D ocean circulation driven by thermal, evaporative, and wind forcings on tidally locked lava planets, providing the first fully 3D framework for magma-ocean dynamics. This framework allows us to characterize circulation patterns and basin structure, derive scaling laws for ocean current speed and ocean depth, and assess the detectability of lava OHT. 

Our simulations reveal a Matsuno–Gill–type response near the equator and geostrophic flows at midlatitudes. Under thermal and evaporative forcings, the flow remains weak and the response is confined within $\pm10^{\circ}$ latitude, set by the equatorial Rossby deformation radius $\sqrt{c_g/\beta}$ ($c_g=\sqrt{\Delta b D_o}\sim20~{\rm m\,s^{-1}}$). When nonlinear advection dominates, the response width instead scales with $\sqrt{U/\beta}$. Wind forcing drives currents up to $\sim100~{\rm m\,s^{-1}}$ in the control case, two orders of magnitude stronger than in the other cases and exceeding the gravity-wave phase speed $c_g$, pushing the Matsuno–Gill response to the edge of the magma ocean.


Based on this framework, we derive scaling laws for ocean current speed $U$, ocean depth $D_o$, and OHT divergence $H$ (Table~\ref{tab:scaling}) and validate them against numerical simulations. Increasing the forcing amplitude enhances both $U$ and $H$, while ocean depth remains below $\sim2000$ m due to stratification, consistent with previous studies \citep{kite2016atmosphere,lai2024a,lai2024b,yang2025} but far shallower than the $\sim100$ km predicted under vigorous internally driven convection \citep[e.g.,][]{leger2011extreme,boukare2022deep}. Magma-ocean heat redistribution $H$ is typically 1–3 orders of magnitude smaller than the stellar flux $S_0$. 
Defining a detectability index $\mathcal{P} \equiv H/S_0$, we combine its scaling with 3D simulations and show that magma-ocean circulation alone cannot produce an observable hotspot offset. Although OHT can be substantial, basin geometry and circulation structure limit the efficiency of longitudinal heat redistribution.


\section*{Acknowledgments}
We are grateful to Shuang Wang, Yixiao Zhang, Yaoxuan Zeng, and Kuan Li for their insightful discussions and constructive feedback. 
Y.L. is supported by the National Natural Science Foundation of China (NSFC) under grant No. 42508019. W.K. is supported by the MIT startup fund. X.T. is supported by NSFC under grant No. 42475131. J.Y. is supported by NSFC under grant No. 42161144011.

\section*{Author contributions:}
Y.L., W.K., and J.Y. designed research; Y.L. performed research; Y.L. and W.K. wrote the manuscript; J.Y. and X.T. improved the manuscript; and all authors discussed the results.

\section*{Data availability:}
The code used to perform simulations with the MITgcm and simulation data used to generate the figures in this study, are stored and accessible at https://zenodo.org/records/18617100.

\appendix
\renewcommand\theequation{\Alph{section}\arabic{equation}}
\renewcommand\thefigure{\Alph{section}\arabic{figure}}
\renewcommand\thetable{\Alph{section}\arabic{table}}
\setcounter{figure}{0}
\setcounter{table}{0}
\setcounter{equation}{0}

\section{Model setup}\label{sec:materials}

\subsection{Governing equations and oceanic parameters}\label{si-sec:model}
We simulate the 3D ocean circulation on tidally locked lava worlds by modifying the MITgcm to solve the primitive equations in spherical coordinates with a nonlinear free surface \citep{adcroft2004}, written as
\begin{equation}
    \frac{D u}{D t} - \frac{u v} {a} tan\phi - f v + \frac{1} {\rho_c a cos \phi} \frac{\partial p'} {\partial \lambda} - {{\nabla}_h \cdot (A_h {\nabla}_h u )} - \frac{\partial} {\partial z} (A_z \frac{\partial u} {\partial z}) = F_{\tau_x} + F_{dx},
\label{equat1}
\end{equation}
\begin{equation}
    \frac{D v}{D t} + \frac{u^2} {a} tan\phi +fu+ \frac{1} {\rho_c a} \frac{\partial p'} {\partial \phi}-{{\nabla}_h \cdot (A_h {\nabla}_h v )} - \frac{\partial} {\partial z} (A_z \frac{\partial v} {\partial z})
     = F_{\tau_y} + F_{dy},
\label{equat2}    
\end{equation}
\begin{equation}
    \frac{\partial \eta}{\partial t} + \frac{1}{acos\phi} \left( \frac{H\hat{u}}{\partial \lambda} + \frac{H\hat{v} cos\phi}{\partial \phi} \right) = F_m,
\label{equat3}    
\end{equation}
\begin{equation}
    \frac{1}{acos \phi} \frac{\partial u} {\partial \lambda} + \frac{1}{a} \frac{\partial v} {\partial \phi} + \frac{\partial w}{\partial z} = 0,
\label{equat4}    
\end{equation}
\begin{equation}
    \rho=\rho(T,p),
\label{eq:eos}
\end{equation} 
\begin{equation}
    p' = \rho_c g\eta + \int_z^0 \rho' g dz,
\label{equat5}
\end{equation} 
\begin{equation}
    \frac{D \theta}{D t}- {{\nabla}_h \cdot (k_h {\nabla}_h \theta )} - 
    \frac{\partial}{\partial z} (k_z \frac{\partial \theta}{\partial z} )= F_{\theta}, 
\label{equat6}    
\end{equation}
where $\lambda$, $\phi$, and $z$ are longitude, latitude, and vertical distance from the surface (negative); $u$, $v$, and $w$ are zonal, meridional, and vertical velocities, respectively; $\theta$ and $T$ are potential temperature and temperature ($\theta \approx T$ within a shallow domain); $\rho$ is silicate density, and $\rho_c$ = 2600 kg\,m$^{-3}$ is the reference density of molten silicates, $\rho' = \rho-\rho_c$ is the density anomaly; $\eta$ is sea surface height (SSH), determined by the vertical integral of horizontal divergence of horizontal velocity, $H\hat{u} = \int_{-H}^\eta u dz$ and $H\hat{v} = \int_{-H}^\eta v dz$, where $H$ is the depth of the simulated domain; $a$ is the planetary radius; $p'$ is the pressure anomaly, composed of a barotropic part due to variations in surface height and a hydrostatic part due to the vertical integral of density anomaly; $\frac{D}{Dt} = \frac{\partial} {\partial t}+ \frac{1}{acos \phi} \frac{\partial} {\partial \lambda} + \frac{1}{a} \frac{\partial} {\partial \phi} + w\frac{\partial}{\partial z}$ is the total derivative; $g$ is gravity; $f = 2 \Omega sin \phi$ is the Coriolis parameter, where $\Omega$ is the planetary rotation rate, $A_h$ and $A_z$ are horizontal and vertical viscosity coefficients, $k_h$ and $k_v$ are horizontal and vertical diffusivities for potential temperature. 

Multicomponent silicates start to melt at the solidus (T$_{\rm sol}$) and become fully molten at the liquidus (T$_{\rm liq}$) \citep{zilinskas2022observability}. Both the solidus and the liquidus dependent on pressure \citep{monteux2016cooling,zhang2022internal}. Given the relatively shallow domain considered in this study, we fix these temperatures as constants: T$_{\rm sol}$ =  1700 K and T$_{\rm liq}$ =  2000 K \citep{monteux2016cooling}. 

The equation of state (EoS) describes the dependence of silicate density on temperature and pressure (Eq.~\ref{eq:eos}), typically determined using the third-order Birch-Murnaghan EoS \citep{sakamaki2010density,katsura2010adiabatic}. 
Following \cite{lai2024a}, we use linear and quadratic approximations for the temperature and pressure dependencies of density, respectively, derived from the third-order Birch–Murnaghan EoS. The full EoS expressions can be found in \cite{lai2024a}. In brief, density decreases linearly with temperature both below the solidus and above the liquidus, with a thermal expansion coefficient $\alpha=8\times10^{-5}$ K$^{-1}$. Between the solidus and liquidus, density decreases with temperature following a hyperbolic tangent (tanh) profile, with a total density contrast of 10\% \citep{monteux2016cooling}. Density increases quadratically with pressure throughout the layer.   

Ocean circulation is driven by three types of external forcing: thermal ($F_{\theta}$), evaporative ($F_m$), and wind forcings ($F_{\tau_x}$ \& $F_{\tau_y}$). 
The surface temperature is relaxed toward a prescribed radiative equilibrium profile $\theta^\star$ over a radiative timescale $\tau_{\rm rad}$:
\begin{equation}
    F_{\theta}=-\frac{1}{\tau_{\rm rad}} (\theta-\theta^{\star}).
    \label{eq:Ftheta}
\end{equation}
The equilibrium temperature within decays away from the SP following $T_{\rm sub}\cos^{1/4}\lambda\cos^{1/4}\phi$, where $T_{\rm sub}$ is the substellar temperature, with a minimum temperature of 50~K beyond $\pm110^{\circ}$ \citep{leger2009transiting}.
The radiative relaxation timescale can be estimated from
\begin{equation}
    \tau_{\rm rad} = \frac{\rho_c c_p h}{4 \sigma T_{\rm sub}^3}, 
    \label{eq:tau_rad}
\end{equation}
where $c_p = 1800$ J\,kg$^{-1}$\,K$^{-1}$ is the heat capacity of typical molten silicates \citep{monteux2016cooling}, $h = 5$ m is the thickness of the surface layer, $\sigma$ is the Stefan-Boltzmann constant, and $T_{\rm sub} = 3000$ K, yielding $\tau_{\rm rad} \approx 4000$ s.
Evaporative forcing drives ocean circulation through modifying the mass (i.e., SSH), written as
\begin{equation}
    F_m = -\frac{E}{\rho_c}.
    \label{eq:Fm}
\end{equation} 
where $E$ is the evaporation rate.
Zonal and meridional wind forcings are given by,
\begin{equation}
    F_{\tau_x}= \frac{\tau_x} {\rho_c h }, F_{\tau_y} = \frac{\tau_y}{\rho_c h},
    \label{eq:Ftaux}
\end{equation}
where $\tau_x$ and $\tau_y$ are zonal and meridional wind stresses.
Note that external forcings are applied only in the surface layer of the model.

To disable dynamics and motions in non-liquid regions where temperatures lie below the liquidus, we apply a strong horizontal drag with a relaxation timescale $\tau_{\rm drag}$ = 20 s for $T<T_{\rm liq}$, rather than imposing an extremely high viscosity \citep{zhang2022internal,lai2024a}.
\begin{equation}
    F_{dx}=-\frac{u}{\tau_{\rm drag}}, \quad 
    F_{dy}=-\frac{v}{\tau_{\rm drag}} \quad (T<T_{liq}).
    \label{eq:rbc}
\end{equation}
Temperature advection is further disabled in non-liquid regions to avoid unphysical heat transport.

\begin{table}[htbp]
\caption{Basic Parameters Used in the Simulations\label{si-table:paras}}
\centering
\small
\begin{tabular}{lll@{\hspace{3em}}lll}
\hline
Parameter & Value (Unit) &  & Parameter & Value (Unit) &  \\
\hline
Planetary radius ($a$) & 9000 km &  & Heat capacity ($c_p$) & 1800 J\,kg$^{-1}$\,K$^{-1}$ &  \\
Surface gravity ($g$) & 22 m\,s$^{-2}$ &  & Vertical viscosity ($A_z$) & $10^{-4}$ m$^2$\,s$^{-1}$ &  \\
Planetary rotation rate ($\Omega$) & $9\times10^{-5}$ s$^{-1}$ &  & Horizontal diffusivity ($k_h$) & $10^{-3}$ m$^2$\,s$^{-1}$ &  \\
Thermal expansion coeff. ($\alpha$) & $8\times10^{-5}$ K$^{-1}$ &  & Vertical diffusivity ($k_z$) & $10^{-4}$ m$^2$\,s$^{-1}$ &  \\
Vertical depth & 4000 m &  & Solidus temperature ($T_{\rm sol}$) & 1700 K &  \\
Density contrast ($\Delta\rho$) & 10\% &  & Liquidus temperature ($T_{\rm liq}$) & 2000 K &  \\
Temp. relaxation timescale ($\tau_{\theta}$) & 4000/1000 s  &  & Drag timescale ($\tau_{\rm rbc}$) & 20 s &   \\
Horizontal resolution & 160$\times$72/480$\times$216  &  & Vertical resolution & 44/176 layers &   \\
\hline
\end{tabular}
\end{table}

Silicate viscosity changes significantly with melt fraction, ranging from $10^{-4}$ m$^2$\,s$^{-1}$ at the liquidus to $10^{18}$ m$^2$\,s$^{-1}$ at the solidus \citep{solomatov2007magma,zhang2022internal}, a range far beyond the resolvable range of current GCMs. For high-temperature molten silicates, viscosity can be approximated as a constant \citep{zhang2022internal}. 
Because motions in non-liquid regions are already suppressed by the imposed drag, we use a uniform vertical viscosity $A_z = 10^{-4}$ m$^2$\,s$^{-1}$, comparable to the molecular viscosity of molten silicates. Horizontal viscosity follows the Leith viscosity scheme \citep{leith1996}, which parameterizes subgrid-scale turbulence and increases with flow deformation, reaching values up to $10^7$ m$^2$\,s$^{-1}$ in regions where velocities exceed 100 m\,s$^{-1}$.
We set the horizontal and vertical diffusivities to $k_h = 10^{-3}$ and $k_z = 10^{-4}$ m$^2$\,s$^{-1}$, similar to eddy diffusivities in Earth’s oceans \citep{Waterhouse2014}. 

To ensure the model captures the full evolution of magma ocean depth, the vertical domain depth $H$ must be sufficiently deep to encompass the entire magma ocean at all times. In this study, we adopt a vertical domain depth $H=4$ km, as scaling analyses indicate that magma oceans on lava planets are typically shallower than 2 km \citep{kite2016atmosphere,lai2024b}. Sensitivity tests with varying domain depths further confirm that our results are insensitive to the choice of domain depth $H$, provided that the vertical domain exceeds the maximum expected magma ocean depth \citep{yang2025}.

\subsection{Numerical setup}
The key planetary and oceanic parameters used in the simulations are summarized in Table~\ref{si-table:paras}.
We first perform simulations at a relatively coarse horizontal resolution of 2.25$^{\circ}$ (160$\times$72) with 44 unequally spaced vertical layers. After the system reaches a statistically steady state (Fig.~\ref{si-fig:initial_condition}), the model is restarted from this state at a higher horizontal resolution of 0.75$^{\circ}$ (480$\times$216) with 176 unequally spaced vertical layers. The high-resolution runs are then integrated until a new quasi-equilibrium is achieved. 
In the high-resolution simulations, the radiative timescale is reduced to 1000~s, due to the reduction in surface layer thickness from 5~m to 1.25~m (Eq.~\ref{eq:tau_rad}).

Simulations are initialized from rest and a cold state, with temperatures below the solidus everywhere in the domain. As shown in the Supplementary Text, the solution is insensitive to the initial temperature conditions (Fig.~\ref{si-fig:initial_condition}). 
A no-flux boundary condition for temperature is applied at the bottom boundary.

\setcounter{figure}{0}
\setcounter{equation}{0}
\section{Robustness to Initial Temperature Conditions}\label{si-sec:ini_con}
\begin{figure}[t]
    \centering
    \includegraphics[width=0.8\textwidth]{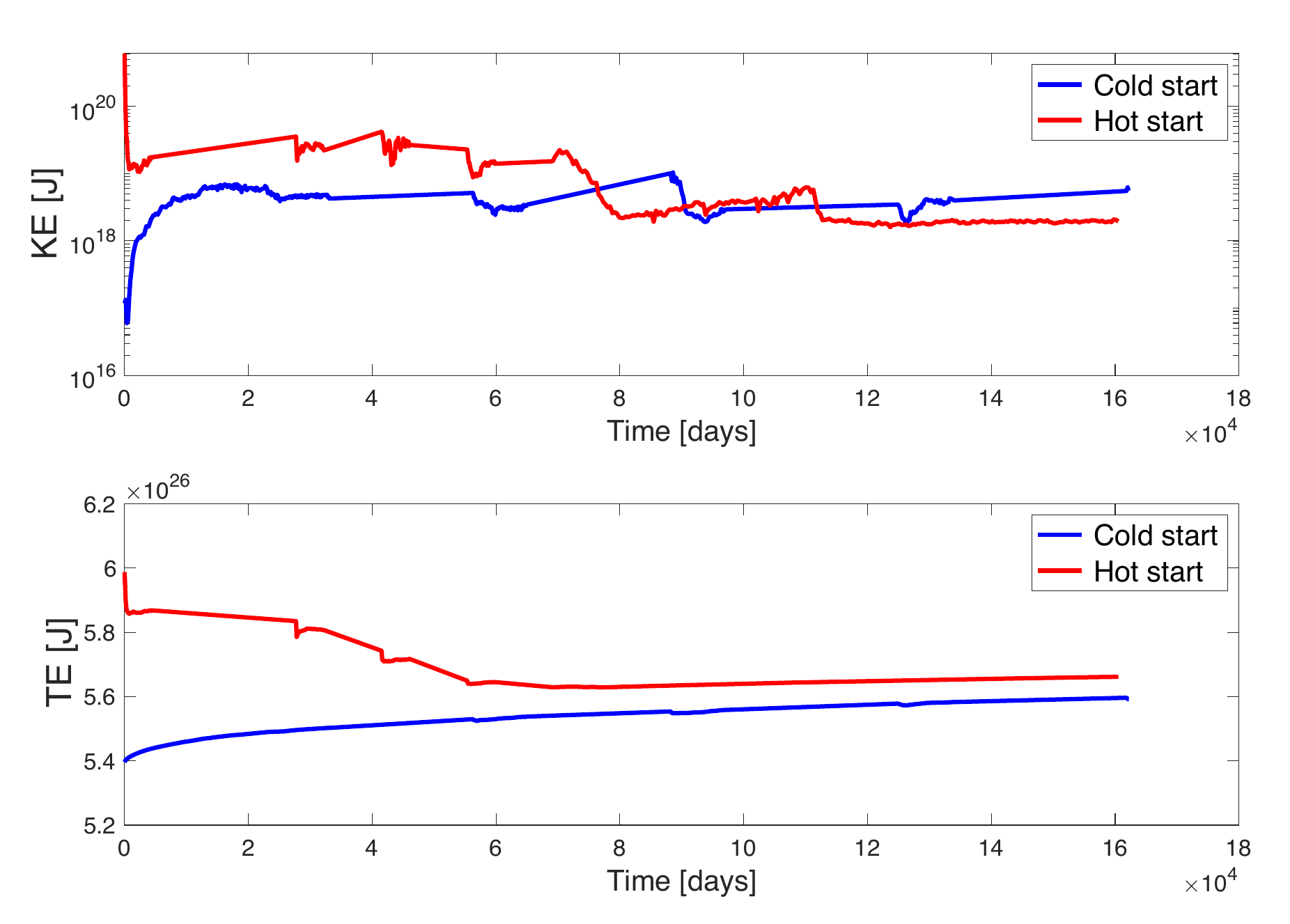}
        \caption{Robustness to initial temperature states. Time series of total kinetic energy (top panel) and total energy (bottom panel) from simulations initialized with two different temperature states: a cold start (blue lines) and a hot start (red lines). The total kinetic energy is defined as KE$=\frac{1}{2} \int_{-H}^0 \int_{-\pi}^{\pi} \int_0^{2\pi} \rho_c a^2 \cos\phi \, (u^2+v^2) \, d\lambda \, d\phi \, dz$, and the total energy as TE$=\int_{-H}^0 \int_{-\pi}^{\pi} \int_0^{2\pi} \rho_c a^2 \cos\phi \, c_p T \, d\lambda \, d\phi \, dz$, where $\lambda$, $\phi$, and $z$ are longitude, latitude, and depth, respectively; $u$, $v$, and $T$ are the zonal velocity, meridional velocity, and temperature; $a$ is the planetary radius; $c_p$ is the specific heat capacity; $\rho_c$ is the silicate density; and $H$ is the vertical extent of the model domain. The discontinuity in the curves results from an artificial acceleration of temperature evolution used to speed up convergence.}
    \label{si-fig:initial_condition}
\end{figure}

To assess the sensitivity of the equilibrium solution to initial thermal conditions, we perform two wind-driven simulations initialized from contrasting states: a molten hot start and a fully solid cold start.
In the hot-start case, the interior temperature is initially vertically uniform and equal to the local surface temperature, such that regions exceeding the liquidus are molten down to the model bottom. In the cold-start case, the temperature is everywhere below the solidus, representing a fully solid planet.

As shown in Fig.~\ref{si-fig:initial_condition}, both simulations converge to the same statistically steady state in terms of total kinetic energy and total energy, demonstrating that the final solution is independent of the initial thermal structure.
In the cold-start case, downward heat transport by vertical diffusion and advection progressively deepens the magma ocean, leading to a gradual increase in both kinetic and total energy. In contrast, the hot-start case initially exhibits strong horizontal currents driven by large temperature gradients throughout the vertical column, which rapidly reorganize the interior structure and produce a weakly stratified deep ocean.

To accelerate equilibration in the hot-start configuration, we apply an iterative procedure combining natural integration with temporally extrapolated thermal adjustment in the solid regions. This approach reduces the long spin-up timescale without altering the final equilibrium state. The discontinuities in Fig.~\ref{si-fig:initial_condition} correspond to these acceleration steps. As the magma ocean shoals, kinetic and total energy decrease and ultimately converge to the cold-start solution.

\setcounter{figure}{0}
\setcounter{equation}{0}
\section{Momentum and heat budget}\label{si-sec:mom-thermo}

\begin{figure}
    \centering
    \includegraphics[width=0.98\textwidth]{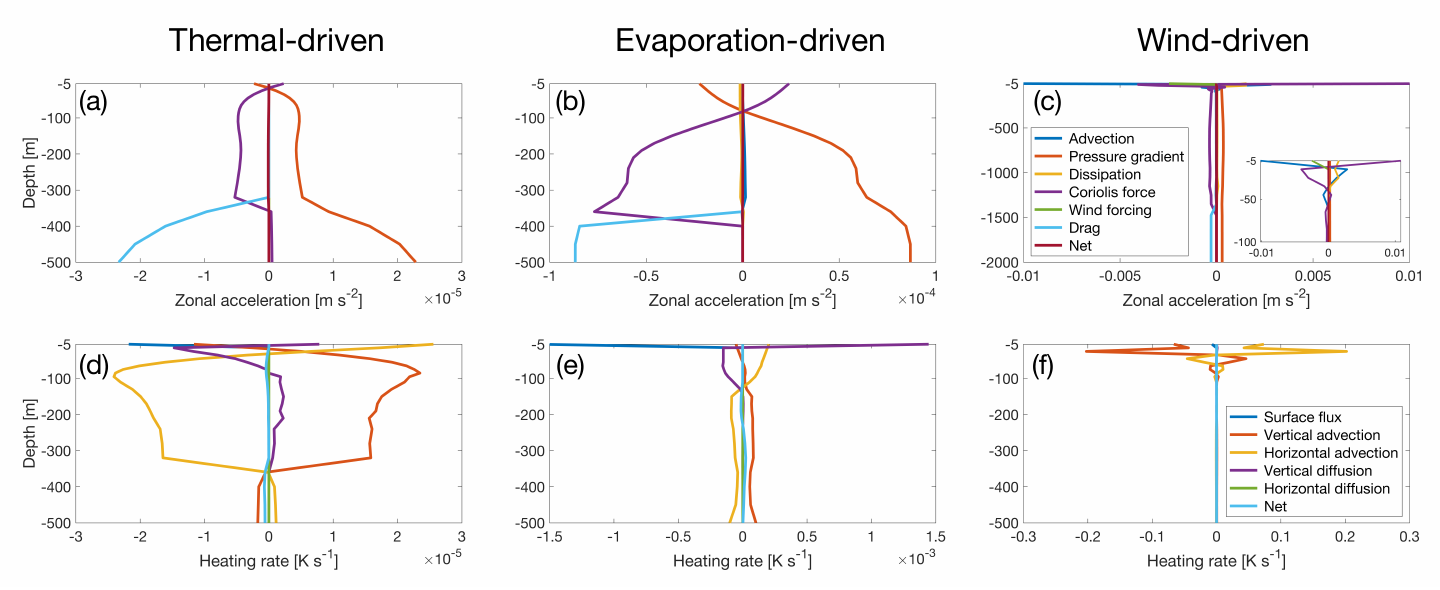}
    \caption{Zonal momentum (top) and heat budget (bottom) in the thermal-driven (left), evaporation-driven (middle), and wind-driven (right) cases. Top: zonal momentum budget (m\,s$^{-2}$) at (50$^{\circ}$W, 30$^{\circ}$N) as a function of depth. The inset in panel (c) shows a magnified view in the upper 100 m. Bottom: heat budget (K\,s$^{-1}$) at (50$^{\circ}$W, 30$^{\circ}$N) as a function of depth. Note that the presented vertical range is different among the three cases. 
    }
    \label{si-fig:mom-thermo}
\end{figure}

Fig.~\ref{si-fig:mom-thermo}(a-c) illustrates the zonal momentum budget of different cases. The thermal- and evaporation-driven cases show similar horizontal momentum structures (Fig.~\ref{si-fig:mom-thermo}(a-b)). Near the equator, the zonal momentum is maintained by a balance among the pressure gradient force, advection, dissipation, and the Coriolis force. Away from the equator, the flow within the ocean is in geostrophic balance between the pressure gradient force and the Coriolis force. Outside the magma ocean, the pressure gradient force is balanced by the imposed drag in non-liquid regions, which effectively damps the horizontal velocity toward zero. When wind forcing is apllied, the ocean interior remains in geostrophic balance (Fig.~\ref{si-fig:mom-thermo}(c)). However, near the surface, the nonlinear advection term becomes comparable to the Coriolis force and wind forcing (the inset in Fig.~\ref{si-fig:mom-thermo}(c)).

The heat budget closes well in all three cases, both at the surface and in the interior ocean (Fig.~\ref{si-fig:mom-thermo}(d-f)). Regardless of the type of external forcing, horizontal and vertical advection dominate the heat budget. In the thermal- and evaporation-driven cases, vertical diffusion also contributes significantly, especially near the surface. In the wind-driven case, the heating rate below 100 m is much smaller than in the upper layers, due to the rapid decrease of current speed with depth. 

\setcounter{figure}{0}
\setcounter{equation}{0}
\section{Thermal phase curve}\label{si-sec:fpfs}

\begin{figure}
    \centering
    \includegraphics[width=0.8\columnwidth]{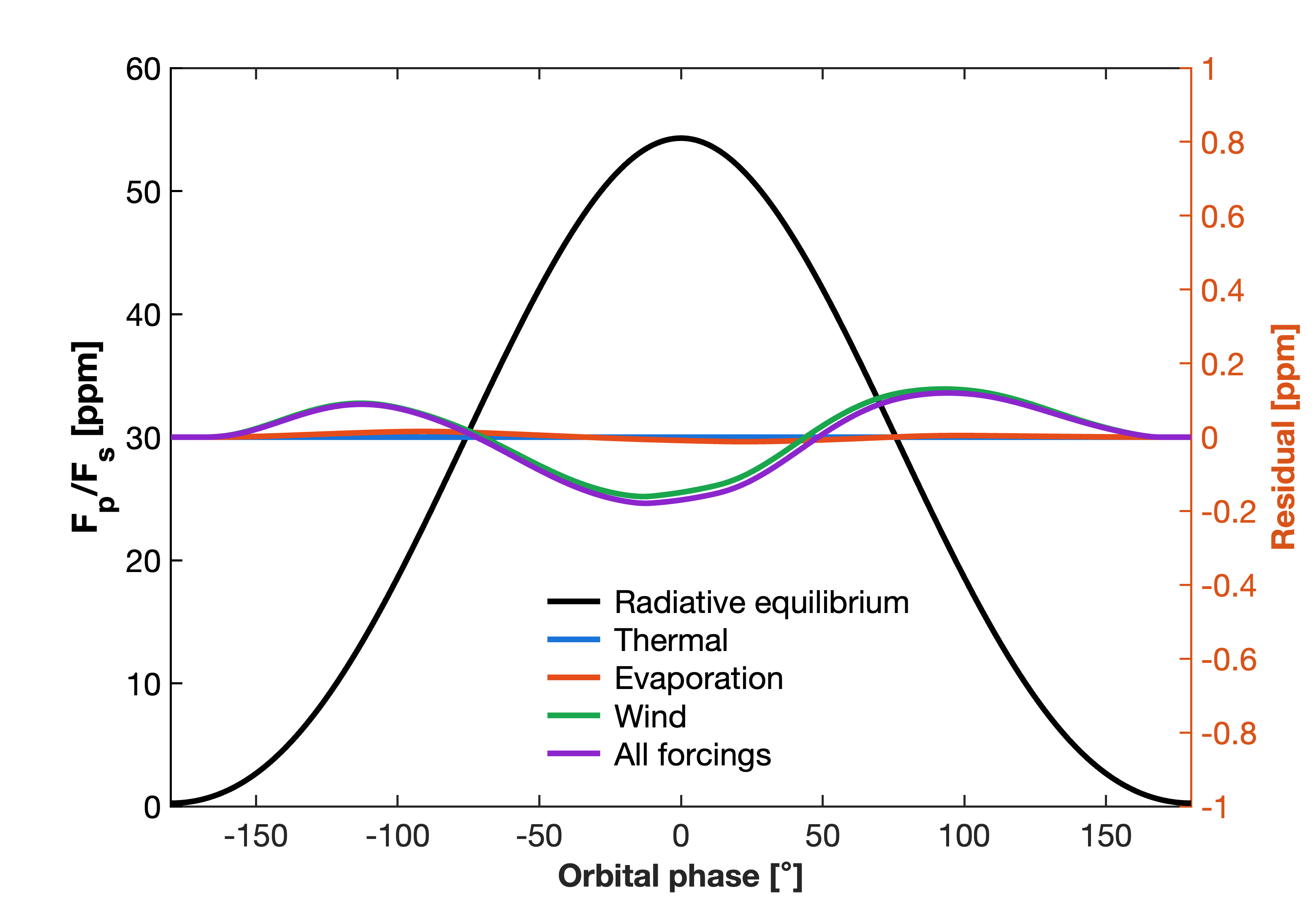}
    \caption{Planet-to-star flux ratio as a function of orbital phase for Kepler-10 b. The black curve represents radiative equilibrium (no circulation), while the colored curves represent residual flux ratios relative to the radiative equilibrium for thermal-driven (blue), evaporation-driven (orange), wind-driven (green), and all forcings (purple) cases. The blue and orange curves nearly overlap, while the green and purple curves nearly overlap.}
    \label{si-fig:fpfs}
\end{figure}

We compute the thermal phase curves of tidally locked lava planets following the approach of \citep{maurin2012, selsis2013}. The thermal phase curve, defined as the planet-to-star flux ratio as a function of orbital phase, is obtained from the ratio between the disk-integrated bolometric emission $F_p$ and the bolometric stellar flux $F_{\rm star}$, received at the orbital distance $d$.

Assuming blackbody emission, the disk-integrated bolometric thermal flux of the planet at distance $d$ is
\begin{equation}
    F_p  = \sum_j I_j (T_j) \frac{S_j cos(\alpha_j)}{d^2}, 
    \label{eq:Fp}
\end{equation}
where $S_j$ is the surface area of the cell $j$, $\alpha_j$ is the angle between the surface normal and the direction toward the observer, $I_j=\epsilon_{\lambda} B_{\lambda}(T_j)$ is the specific intensity at surface temperature $T_j$, $\epsilon_{\lambda}=1$ is the surface emissivity, and $B_{\lambda}$ is the Plank function. The geometric factor $cos(\alpha_j)$ is given by
\begin{equation}
    \mu = \rm cos(\alpha_j) = \rm cos(lat_j-lat_{obs})\cdot cos(lon_j-lon_{obs}), 
    \label{eq:Fp_mu}
\end{equation}
where lon$_j$ and lat$_j$ are the longitude and latitude of the cell $j$, and lon$_{\rm obs}$ and lat$_{\rm obs}$ are the longitude and latitude of the sub-observer point. For plants with zero obliquity, lat$_{\rm obs}$ = 0, while lon$_{\rm obs}$ varies from 0 to 2$\pi$. Only grid cells visible to the observer ($cos(\alpha_j)>0$) contribute to $F_p$. 

The bolometric stellar flux at distance $d$ is
\begin{equation}
    F_{\rm star}  = \pi  B_{\lambda} (T_{\rm star}) (\frac{R_{star}}{d})^2, 
    \label{eq:Fs}
\end{equation}
where $T_{\rm star}$ and $R_{\rm star}$ are the stellar effective temperature and radius. The wavelength-dependent thermal phase curve is therefore quantified by $F_p/F_{\rm star}$. In this study, we take the band average values between 5 and 20 $\mu$m.

Fig.~\ref{si-fig:fpfs} shows the simulated thermal phase curves adopting Kepler-10 b's stellar parameters, calculated from the disk-integrated bolometric planetary emission normalized by the stellar flux. All cases produce similar phase curves, with planet-to-star flux ratios peaking at $\sim$55~ppm at zero orbital phase and exhibiting no phase offset, confirming the limited influence of magma-ocean circulation on the thermal phase curve for Kepler-10 b. Moreover, the differences in flux ratios among the simulations remain below 1~ppm, well below the detection threshold of current facilities, including JWST \citep{kempton2024}.

\bibliography{sample701}{}

@article{kang2021escaping,
  title={Escaping outflows from disintegrating exoplanets: day-side versus night-side escape},
  author={Kang, Wanying and Ding, Feng and Wordsworth, Robin and Seager, Sara},
  journal={The Astrophysical Journal},
  volume={906},
  number={2},
  pages={67},
  year={2021},
  doi={10.3847/1538-4357/abcaa7},
  publisher={IOP Publishing}
}

@article{yang2025,
  title={Ocean Circulation on Tide-locked Lava Worlds: 3D Modeling with a Simple Boundary Iteration Method},
  author={Yang, Jun and Tang, Chengyao and Wang, Zimu and Lai, Yanhong and Kang, Wanying},
  journal={The Astrophysical Journal Letters},
  volume={995},
  number={1},
  pages={L20},
  year={2025},
  DOI={10.3847/2041-8213/ae1cc0},
  publisher={IOP Publishing}
}

@article{Rhines_1975, 
    title={Waves and turbulence on a beta-plane}, 
    volume={69}, 
    DOI={10.1017/S0022112075001504}, 
    number={3}, 
    journal={Journal of Fluid Mechanics}, 
    author={Rhines, Peter B.}, 
    year={1975}, 
    pages={417–443}}

@article{kang2023true,
  title={True Polar Wander of Lava Worlds},
  author={Kang, Wanying and Nimmo, Francis and Ding, Feng},
  journal={The Astrophysical Journal Letters},
  volume={949},
  number={2},
  pages={L20},
  year={2023},
  doi={10.3847/2041-8213/acd691},
  publisher={IOP Publishing}
}

@article{kempton2024,
  title={Transiting exoplanet atmospheres in the era of JWST},
  author={Kempton, Eliza M-R and Knutson, Heather A},
  journal={Reviews in Mineralogy and Geochemistry},
  volume={90},
  number={1},
  pages={411--464},
  year={2024},
  url={https://doi.org/10.2138/rmg.2024.90.12},
  publisher={Mineralogical Society of America}
}

@article{zeng2021,
  title={Oceanic superrotation on tidally locked planets},
  author={Zeng, Yaoxuan and Yang, Jun},
  journal={The Astrophysical Journal},
  volume={909},
  number={2},
  pages={172},
  year={2021},
  DOI={10.3847/1538-4357/abe12f},
  publisher={IOP Publishing}
}

@article{hu2014,
  title={Role of ocean heat transport in climates of tidally locked exoplanets around M dwarf stars},
  author={Hu, Yongyun and Yang, Jun},
  journal={Proceedings of the National Academy of Sciences},
  volume={111},
  number={2},
  pages={629--634},
  year={2014},
  url={https://doi.org/10.1073/pnas.1315215111},
  publisher={National Academy of Sciences}
}

@article{gill1980,
  title={Some simple solutions for heat-induced tropical circulation},
  author={Gill, Adrian E},
  journal={Quarterly Journal of the Royal Meteorological Society},
  volume={106},
  number={449},
  pages={447--462},
  year={1980},
  publisher={Wiley Online Library}
}

@article{matsuno1966,
  title={Quasi-geostrophic motions in the equatorial area},
  author={Matsuno, Taroh},
  journal={Journal of the Meteorological Society of Japan. Ser. II},
  volume={44},
  number={1},
  pages={25--43},
  year={1966},
  publisher={Meteorological Society of Japan}
}

@inproceedings{adcroft2004,
  title={Overview of the formulation and numerics of the MIT GCM},
  author={Adcroft, Alistair and Hill, Chris and Campin, Jean-Michel and Marshall, John and Heimbach, Patrick},
  booktitle={Proceedings of the ECMWF seminar series on Numerical Methods, Recent developments in numerical methods for atmosphere and ocean modelling},
  pages={139--149},
  year={2004}
}

@article{leger2009transiting,
  title={Transiting exoplanets from the CoRoT space mission-VIII. CoRoT-7b: The first super-Earth with measured radius},
  author={L{\'e}ger, A and Rouan, D and Schneider, Jodi and Barge, P and Fridlund, M and Samuel, B and Ollivier, M and Guenther, E and Deleuil, M and Deeg, HJ and others},
  journal={Astronomy \& Astrophysics},
  volume={506},
  number={1},
  pages={287--302},
  year={2009},
  publisher={EDP Sciences},
  DOI={10.1051/0004-6361/200911933}
}

@article{leger2011extreme,
  title={The extreme physical properties of the CoRoT-7b super-Earth},
  author={L{\'e}ger, Alain and Grasset, O and Fegley, B and Codron, F and Albarede, AF and Barge, P and Barnes, R and Cance, P and Carpy, S and Catalano, F and others},
  journal={Icarus},
  volume={213},
  number={1},
  pages={1--11},
  year={2011},
  publisher={Elsevier},
  doi={doi:10.1016/j.icarus.2011.02.004}
}

@article{rouan2011,
  title={The orbital phases and secondary transits of Kepler-10b. A physical interpretation based on the lava-ocean planet model},
  author={Rouan, Daniel and Deeg, Hans J and Demangeon, Olivier and Samuel, Benjamin and Cavarroc, C{\'e}line and Fegley, Bruce and L{\'e}ger, Alain},
  journal={The Astrophysical Journal Letters},
  volume={741},
  number={2},
  pages={L30},
  year={2011},
  DOI={10.1088/2041-8205/741/2/L30},
  publisher={IOP Publishing}
}

@article{batalha2011kepler,
  title={Kepler's first rocky planet: Kepler-10b},
  author={Batalha, Natalie M and Borucki, William J and Bryson, Stephen T and Buchhave, Lars A and Caldwell, Douglas A and Christensen-Dalsgaard, J{\o}rgen and Ciardi, David and Dunham, Edward W and Fressin, Francois and Gautier, Thomas N and others},
  journal={The Astrophysical Journal},
  volume={729},
  number={1},
  pages={27},
  year={2011},
  publisher={IOP Publishing},
  doi={10.1088/0004-637X/729/1/27}
}

@article{selsis2013,
  title={The effect of rotation and tidal heating on the thermal lightcurves of super Mercuries},
  author={Selsis, Franck and Maurin, A-S and Hersant, F and Leconte, J and Bolmont, Emeline and Raymond, Sean N and Delbo, M},
  journal={Astronomy \& Astrophysics},
  volume={555},
  pages={A51},
  year={2013},
  DOI={10.1051/0004-6361/201321661},
  publisher={EDP Sciences}
}

@article{showman2010,
author = {Showman, Adam P. and Polvani, Lorenzo M.},
title = {The Matsuno-Gill model and equatorial superrotation},
journal = {Geophysical Research Letters},
volume = {37},
number = {18},
pages = {},
doi = {https://doi.org/10.1029/2010GL044343},
url = {https://agupubs.onlinelibrary.wiley.com/doi/abs/10.1029/2010GL044343},
eprint = {https://agupubs.onlinelibrary.wiley.com/doi/pdf/10.1029/2010GL044343},
year = {2010}
}

@article{imamura2020,
  title={Superrotation in planetary atmospheres},
  author={Imamura, Takeshi and Mitchell, Jonathan and Lebonnois, Sebastien and Kaspi, Yohai and Showman, Adam P and Korablev, Oleg},
  journal={Space Science Reviews},
  volume={216},
  number={5},
  pages={87},
  year={2020},
  url={https://doi.org/10.1007/s11214-020-00703-9},
  publisher={Springer}
}

@article{showman2011,
  title={Equatorial superrotation on tidally locked exoplanets},
  author={Showman, Adam P and Polvani, Lorenzo M},
  journal={The Astrophysical Journal},
  volume={738},
  number={1},
  pages={71},
  year={2011},
  doi={10.1088/0004-637X/738/1/71},
  publisher={IOP Publishing}
}

@article{maurin2012,
  title={Thermal phase curves of nontransiting terrestrial exoplanets-II. Characterizing airless planets},
  author={Maurin, AS and Selsis, Franck and Hersant, F and Belu, A},
  journal={Astronomy \& Astrophysics},
  volume={538},
  pages={A95},
  year={2012},
  DOI={10.1051/0004-6361/201117054},
  publisher={EDP Sciences}
}

@article{stommel1948,
  author = {Stommel, Henry},
  title = {The Westward Intensification of Wind-Driven Ocean Currents},
  journal = {Transactions of the American Geophysical Union},
  year = {1948},
  volume = {29},
  pages = {202--206},
  doi = {10.1029/TR029i002p00202},
}

@article{munk1950,
  author = {Munk, Walter},
  title = {On the Wind-Driven Ocean Circulation},
  journal = {Journal of Meteorology},
  year = {1950},
  volume = {7},
  pages = {80--93},
  doi = {10.1175/1520-0469(1950)007<0080:OTWDOC>2.0.CO;2},
}

@article{bourrier201855,
  title={The 55 Cancri system reassessed},
  author={Bourrier, Vincent and Dumusque, Xavier and Dorn, Caroline and Henry, Gregory W and Astudillo-Defru, Nicola and Rey, Javiera and Benneke, Bj{\"o}rn and H{\'e}brard, Guillaume and Lovis, Christophe and Demory, Brice-Olivier and others},
  journal={Astronomy \& Astrophysics},
  volume={619},
  pages={A1},
  year={2018},
  publisher={EDP Sciences},
  url={https://doi.org/10.1051/0004-6361/201833154}
}

@article{brinkman2023toi,
  title={TOI-561 b: A Low-density Ultra-short-period “Rocky” Planet around a Metal-poor Star},
  author={Brinkman, Casey L and Weiss, Lauren M and Dai, Fei and Huber, Daniel and Kite, Edwin S and Valencia, Diana and Bean, Jacob L and Beard, Corey and Behmard, Aida and Blunt, Sarah and others},
  journal={The Astronomical Journal},
  volume={165},
  number={3},
  pages={88},
  year={2023},
  publisher={IOP Publishing},
  url={https://doi.org/10.3847/1538-3881/acad83}
}

@article{boukare2022deep,
  title={Deep two-phase, hemispherical magma oceans on lava planets},
  author={Boukar{\'e}, Charles-{\'E}douard and Cowan, Nicolas B and Badro, James},
  journal={The Astrophysical Journal},
  volume={936},
  number={2},
  pages={148},
  year={2022},
  publisher={IOP Publishing},
  url={https://doi.org/10.3847/1538-4357/ac8792}
}

@article{monteux2016cooling,
  title={On the cooling of a deep terrestrial magma ocean},
  author={Monteux, Julien and Andrault, Denis and Samuel, Henri},
  journal={Earth and Planetary Science Letters},
  volume={448},
  pages={140--149},
  year={2016},
  publisher={Elsevier},
  url={http://dx.doi.org/10.1016/j.epsl.2016.05.010}
}

@article{kite2016atmosphere,
  title={Atmosphere-interior exchange on hot, rocky exoplanets},
  author={Kite, Edwin S and Fegley Jr, Bruce and Schaefer, Laura and Gaidos, Eric},
  journal={The Astrophysical Journal},
  volume={828},
  number={2},
  pages={80},
  year={2016},
  publisher={IOP Publishing},
  doi={10.3847/0004-637X/828/2/80}
}

@article{malavolta2018ultra,
  title={An ultra-short period rocky super-Earth with a secondary eclipse and a Neptune-like companion around K2-141},
  author={Malavolta, Luca and Mayo, Andrew W and Louden, Tom and Rajpaul, Vinesh M and Bonomo, Aldo S and Buchhave, Lars A and Kreidberg, Laura and Kristiansen, Martti H and Lopez-Morales, Mercedes and Mortier, Annelies and others},
  journal={The Astronomical Journal},
  volume={155},
  number={3},
  pages={107},
  year={2018},
  publisher={IOP Publishing},
  url={https://doi.org/10.3847/1538-3881/aaa5b5}
}

@article{castan2011atmospheres,
  title={Atmospheres of hot super-Earths},
  author={Castan, Thibaut and Menou, Kristen},
  journal={The Astrophysical Journal Letters},
  volume={743},
  number={2},
  pages={L36},
  year={2011},
  publisher={IOP Publishing},
  doi={10.1088/2041-8205/743/2/L36}
}

@article{nguyen2020modelling,
  title={Modelling the atmosphere of lava planet K2-141b: implications for low-and high-resolution spectroscopy},
  author={Nguyen, T Giang and Cowan, Nicolas B and Banerjee, Agnibha and Moores, John E},
  journal={Monthly Notices of the Royal Astronomical Society},
  volume={499},
  number={4},
  pages={4605--4612},
  year={2020},
  publisher={Oxford University Press},
  doi={10.1093/mnras/staa2487}
}

@article{nguyen2022impact,
  title={The impact of ultraviolet heating and cooling on the dynamics and observability of lava planet atmospheres},
  author={Nguyen, T Giang and Cowan, Nicolas B and Pierrehumbert, Raymond T and Lupu, Roxana E and Moores, John E},
  journal={Monthly Notices of the Royal Astronomical Society},
  volume={513},
  number={4},
  pages={6125--6133},
  year={2022},
  publisher={Oxford University Press},
  url={https://doi.org/10.1093/mnras/stac1331}
}

@article{zilinskas2022observability,
  title={Observability of evaporating lava worlds},
  author={Zilinskas, Mantas and Van Buchem, CPA and Miguel, Yamila and Louca, Amy and Lupu, Roxana and Zieba, Sebastian and van Westrenen, Wim},
  journal={Astronomy \& Astrophysics},
  volume={661},
  pages={A126},
  year={2022},
  publisher={EDP Sciences},
  url={https://doi.org/10.1051/0004-6361/202142984}
}

@article{demory2016map,
  title={A map of the large day--night temperature gradient of a super-Earth exoplanet},
  author={Demory, Brice-Olivier and Gillon, Michael and De Wit, Julien and Madhusudhan, Nikku and Bolmont, Emeline and Heng, Kevin and Kataria, Tiffany and Lewis, Nikole and Hu, Renyu and Krick, Jessica and others},
  journal={Nature},
  volume={532},
  number={7598},
  pages={207--209},
  year={2016},
  publisher={Nature Publishing Group UK London},
  doi={10.1038/nature17169}
}

@article{hu2015semi,
  title={A semi-analytical model of visible-wavelength phase curves of exoplanets and applications to Kepler-7 b and Kepler-10 b},
  author={Hu, Renyu and Demory, Brice-Olivier and Seager, Sara and Lewis, Nikole and Showman, Adam P},
  journal={The Astrophysical Journal},
  volume={802},
  number={1},
  pages={51},
  year={2015},
  publisher={IOP Publishing},
  doi={10.1088/0004-637X/802/1/51}
}

@article{ingersoll1985supersonic,
  title={Supersonic meteorology of Io: Sublimation-driven flow of SO2},
  author={Ingersoll, Andrew P and Summers, Michael E and Schlipf, Steve G},
  journal={Icarus},
  volume={64},
  number={3},
  pages={375--390},
  year={1985},
  publisher={Elsevier},
  url={https://doi.org/10.1016/0019-1035(85)90062-4}
}

@book{vallis2019essentials,
  title={Essentials of atmospheric and oceanic dynamics},
  author={Vallis, Geoffrey K},
  year={2019},
  publisher={Cambridge university press}
}

@book{vallis2017atmospheric,
  title={Atmospheric and oceanic fluid dynamics},
  author={Vallis, Geoffrey K},
  year={2017},
  publisher={Cambridge University Press}
}

@article {Waterhouse2014,
      author = "Amy F. Waterhouse and Jennifer A. MacKinnon and Jonathan D. Nash and Matthew H. Alford and Eric Kunze and Harper L. Simmons and Kurt L. Polzin and Louis C. St. Laurent and Oliver M. Sun and Robert Pinkel and Lynne D. Talley and Caitlin B. Whalen and Tycho N. Huussen and Glenn S. Carter and Ilker Fer and Stephanie Waterman and Alberto C. Naveira Garabato and Thomas B. Sanford and Craig M. Lee",
      title = "Global Patterns of Diapycnal Mixing from Measurements of the Turbulent Dissipation Rate",
      journal = "Journal of Physical Oceanography",
      year = "2014",
      publisher = "American Meteorological Society",
      address = "Boston MA, USA",
      volume = "44",
      number = "7",
      doi = "10.1175/JPO-D-13-0104.1",
      pages=      "1854 - 1872",
      url = "https://journals.ametsoc.org/view/journals/phoc/44/7/jpo-d-13-0104.1.xml"
}

@article{solomatov2007magma,
  title={Magma oceans and primordial mantle differentiation},
  author={Solomatov, V},
  journal={Evolution of the Earth},
  volume={9},
  pages={91--119},
  year={2007}
}

@article{zhang2022internal,
  title={Internal dynamics of magma ocean and its linkage to atmospheres},
  author={Zhang, Yizhuo and Zhang, Nan and Tian, Meng},
  journal={Acta geochimica},
  volume={41},
  number={4},
  pages={568--591},
  year={2022},
  publisher={Springer},
  url={https://doi.org/10.1007/s11631-021-00514-x}
}

@article{pedlosky1986buoyancy,
  title={The buoyancy and wind-driven ventilated thermocline},
  author={Pedlosky, Joseph},
  journal={Journal of physical oceanography},
  volume={16},
  number={6},
  pages={1077--1087},
  year={1986},
  url={https://doi.org/10.1175/1520-0485(1986)016<1077:TBAWDV>2.0.CO;2}
}

@article{katsura2010adiabatic,
  title={Adiabatic temperature profile in the mantle},
  author={Katsura, Tomoo and Yoneda, Akira and Yamazaki, Daisuke and Yoshino, Takashi and Ito, Eiji},
  journal={Physics of the Earth and Planetary Interiors},
  volume={183},
  number={1-2},
  pages={212--218},
  year={2010},
  publisher={Elsevier},
  doi={10.1016/j.pepi.2010.07.001}
}

@article{sakamaki2010density,
  title={Density of dry peridotite magma at high pressure using an X-ray absorption method},
  author={Sakamaki, Tatsuya and Ohtani, Eiji and Urakawa, Satoru and Suzuki, Akio and Katayama, Yoshinori},
  journal={American Mineralogist},
  volume={95},
  number={1},
  pages={144--147},
  year={2010},
  publisher={Mineralogical Society of America},
  url={https://doi.org/10.2138/am.2010.3143}
}

@article{sterl2017drag,
  title={Drag at high wind velocities-a review},
  author={Sterl, Andreas},
  journal={Roy. Netherland Met. Office. Tech. Rep},
  volume={361},
  pages={23},
  year={2017},
}

@article{lai2024a,
  title={Ocean Circulation on Tide-locked Lava Worlds. I. An Idealized 2D Numerical Model},
  author={Lai, Yanhong and Yang, Jun and Kang, Wanying},
  journal={The Planetary Science Journal},
  volume={5},
  number={9},
  pages={204},
  year={2024{\noopsort{a}}},
  sortkey={a},
  url={https://doi.org/10.3847/PSJ/ad7111},
  publisher={IOP Publishing}
}

@article{lai2024b,
  title={Ocean Circulation on Tide-locked Lava Worlds. II. Scalings},
  author={Lai, Yanhong and Kang, Wanying and Yang, Jun},
  journal={The Planetary Science Journal},
  volume={5},
  number={9},
  pages={205},
  year={2024{\noopsort{b}}},
  sortkey={b},
  url={https://doi.org/10.3847/PSJ/ad70b4},
  publisher={IOP Publishing}
}

@article{leith1996,
  title={Stochastic models of chaotic systems},
  author={Leith, CE},
  journal={Physica D: Nonlinear Phenomena},
  volume={98},
  number={2-4},
  pages={481--491},
  year={1996},
  url={https://doi.org/10.1016/0167-2789(96)00107-8},
  publisher={Elsevier}
}

@article{hu2024,
  title={A secondary atmosphere on the rocky exoplanet 55 Cancri e},
  author={Hu, Renyu and Bello-Arufe, Aaron and Zhang, Michael and Paragas, Kimberly and Zilinskas, Mantas and van Buchem, Christiaan and Bess, Michael and Patel, Jayshil and Ito, Yuichi and Damiano, Mario and others},
  journal={Nature},
  volume={630},
  number={8017},
  pages={609--612},
  year={2024},
  url={https://doi.org/10.1038/s41586-024-07432-x},
  publisher={Nature Publishing Group UK London}
}

@article{zieba2022,
  title={K2 and Spitzer phase curves of the rocky ultra-short-period planet K2-141 b hint at a tenuous rock vapor atmosphere},
  author={Zieba, Sebastian and Zilinskas, Mantas and Kreidberg, Laura and Nguyen, Tue Giang and Miguel, Yamila and Cowan, Nicolas B and Pierrehumbert, Ray and Carone, Ludmila and Dang, Lisa and Hammond, Mark and others},
  journal={Astronomy \& Astrophysics},
  volume={664},
  pages={A79},
  year={2022},
  url={https://doi.org/10.1051/0004-6361/202142912},
  publisher={EDP Sciences}
}

@article{angelo2017,
  title={A case for an atmosphere on super-Earth 55 Cancri e},
  author={Angelo, Isabel and Hu, Renyu},
  journal={The Astronomical Journal},
  volume={154},
  number={6},
  pages={232},
  year={2017},
  doi={10.1088/0004-637X/729/1/27},
  publisher={IOP Publishing}
}

@article{mercier2022,
  title={Revisiting the iconic spitzer phase curve of 55 Cancri e: hotter dayside, cooler nightside, and smaller phase offset},
  author={Mercier, Samson J and Dang, Lisa and Gass, Alexander and Cowan, Nicolas B and Bell, Taylor J},
  journal={The Astronomical Journal},
  volume={164},
  number={5},
  pages={204},
  year={2022},
  DOI={10.3847/1538-3881/ac8f22},
  publisher={IOP Publishing}
}
\bibliographystyle{aasjournalv7}



\end{document}